\documentclass[twocol]{ametsocV6.1}

\usepackage[greek, english]{babel}
\usepackage[  
    colorlinks=true,
    linkcolor=red
]{hyperref}
\usepackage{academicons}
\definecolor{orcidlogocol}{HTML}{A6CE39}

\title{A Case Study of the Tornadic Supercell in the Province of Pampanga, Philippines \\ (27 May 2024)}

\authors{Generich H. Capuli\href{https://orcid.org/0000-0003-1253-7043}{\includegraphics[scale=0.5]{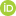}}\aff{a,b}\correspondingauthor{Generich H. Capuli (CGHA), genecap89@gmail.com / genhcapuli@rtu.edu.ph}, Michael Angelo O. Noveno\aff{c}, and Marco Polo A. Ibañez\aff{a}
}
\affiliation{\aff{a}{Research \& Development and Training Division, Department of Science and Technology - Philippine Atmospheric, Geophysical, and Astronomical Services Administration, Brgy. Central, Quezon City 1100, Metro Manila, Philippine}\\
\aff{b}{Project Severe Weather Archive of the Philippines, Quezon City, Philippines}\\
\aff{c}{Albay Public Safety and Emergency Management Office, Old Albay District, Legazpi City 4500, Albay, Philippines}
}

\abstract{This study provides an integrated damage assessment, visual evaluation, environmental context, and remote sensing analysis of the tornado event that struck the suburb of Candating in Arayat, Pampanga on 27 May 2024. Satellite imagery and ground-level damage photographs reveal a $\sim$2 km path, with damage reaching EF2 intensity at one point along the track, based on the Enhanced Fujita (EF) scale. Videos of the tornado and its parent storm reveal a well-defined wall cloud and low-level mesocyclone. Subsequent radar analysis supports these and other features of a tornadic supercell.  Synoptic-scale ascent in the mid- and upper-troposphere was subtle, influenced by the approach of Tropical Cyclone Ewiniar. However, a modest meridional flow aloft provided sufficient deep-layer shear to support supercell development. The southwest monsoon acted as a low-level jet, promoting warm, moist advection into western Luzon. The supercell developed around midday and was characterized by large-undiluted instability, attributed to steep low-level lapse rates. Although low-level shear and the associated near-surface horizontal vorticity were weak, the latter is highly streamwise, allowing for efficient ingestion, tilting, and stretching into vertical vorticity, which aided in tornadogenesis. Both satellite and radar data suggest that storm interactions, such as a nudging mechanism and terrain effects from nearby Mt. Arayat may have contributed to the initiation and intensification of the supercell through lee-side convergence and vorticity enhancement. The complex tropical environment of the Candating, Arayat tornado exhibits several similarities to well-documented tornadic events in North America. These findings highlight the need for further research into the atmospheric conditions conducive to tornadic activity in the Philippines.
}

\begin{document}

\maketitle

%
%
%
%
%
%

%

\section{INTRODUCTION}

Many countries, including the United States and several in Europe, regularly experience a variety of severe convective storm (SCS) hazards that result in the loss of life and property - and the Philippines is no exemption. In 2024 alone, following the launch of a community-based initiative known as the Severe Weather Archive of the Philippines (hereafter Project SWAP), over 200 tornadoes were documented across the country, with approximately 100 occurring over land \citep{Capuli2024b}. Prior to this effort, tornado reports in the country were collected informally at the national level. Most tornadoes observed in the Philippines fall on the weaker end of the Enhanced Fujita (EF) scale (EF0–EF1). Despite their lower intensity, these tornadoes can still cause considerable damage, particularly in densely populated areas. According to \citet{Capuli2024b}, tornadoes occur most frequently in the Greater Metro Manila Region (GMMR) and Central Mindanao, which together account for more than 80\% of reported tornado events nationwide. This geographic concentration increases the potential impact of even weak tornadoes on lives and infrastructures. Notably, these tornadic events are most commonly observed during the Asian Summer Monsoon i.e., Southwest Monsoon (SWM) months of June, July, and August (JJA)\footnote{Locally known as \textit{Hanging Habagat}.}. On the other hand, the United States records a larger proportion of significant to violent tornadoes (EF2 and above), which account for the majority of tornado-related fatalities. These tend to occur during the spring months of March, April, and May (MAM), particularly across the Great Central Plains \citep{Ashley2007,Anderson-Frey2019}. The first formal account of a tornado in the Philippines was conducted by \citet{Capuli2024a} in a thesis that documented the well-observed multiple-vortex Manila Tornado of 14 August 2016, which caused EF1 damage to various structures across Manila City, Metro Manila.

A fundamental framework for characterizing thunderstorm environments involves assessing the presence of key \textit{ingredients} necessary for deep, moist convection. According to \citet{Johns1992}, three essential components are required: (i) sufficient low-level moisture, (ii) conditionally unstable mid-level lapse rates, and (iii) a lifting mechanism capable of raising a potentially buoyant parcel to its level of free convection (LFC). These three allow lift, trigger, and sustain convective updrafts. In addition, for well-organized convective systems such as supercells and squall lines (or quasi-linear convective systems), they depend critically on strong vertical wind shear in the lower to mid-troposphere (typically within the first 6 km). This dynamic ingredient governs storm organization and longevity, as demonstrated in both numerical simulations \citep{Weisman1982} and observational studies \citep{RasmussenBlanchard1998,Thompson2003,Thompson2012,Nixon2022}. \citet{Smith2012} and \citet{Pucik2015} have shown that such organized convective storms are responsible for the majority of significant severe weather events (SWEs) in the United States and Europe, respectively. Tornadoes, in particular, are often linked to the ingestion of low-level streamwise horizontal vorticity aloft (i.e., vorticity along the storm-relative inflow) within the thunderstorms' bases for the updraft to acquire mesocyclonic rotation, the build up and organization of storm outflow and near-surface vertical vorticity, which undergo tilting and stretching into the vertical, ultimately contributing to tornadogenesis \citep{Markowski2003,Nowotarski2013,Parker2014,Coffer2019,Coffer2020,Nixon2022,Fischer2024}. The efficiency of this ‘in-and-up’ process is further enhanced when vertical alignment exists among near-surface ($\sim$500 m), low-level ($\sim$1 km), and mid-level ($\sim$3-6 km) rotation centers \citep{Guarriello2018}.

In the Philippines, the Department of Science and Technology - Philippine Atmospheric, Geophysical, and Astronomical Services Administration (DOST-PAGASA) has colloquially referred to the month of May as the “Thunderstorm Month”. However, recent climatological analyses indicate that May actually signals the beginning of the country's severe weather season \citep{Capuli2024b}. As discussed in earlier sections and highlighted by the Project SWAP, both tornadic supercell (mesocyclonic) and non-supercell (non-mesocyclonic) thunderstorms capable of producing widespread severe weather, including tornadoes, most frequently occur in the GMMR, particularly during SWM period from May to September \citep{Peralta2020}. During this period,  warm and moist low-level winds are advected toward the western section of Luzon, coinciding with increased instability, steep lapse rates, and enhanced vertical wind shear, which together favor supercell development. The months of June through August (JJA) represent the peak of the severe weather season and often align with early-season tropical cyclone (TC) incursions.

Case studies around the world demonstrate how synoptic-scale and mesoscale environments influence tornadogenesis. For instance, \citet{Thompson2000} analyzed the 3 May 1999 event in the U.S. Great Plains, \citet{Oliveira2022} documented the 20 April 2015 event in Brazil, and \citet{Choo2018} examined the 6 May 2012 tornado in Japan. All aforementioned authors concluded that even under weak synoptic forcing (i.e., frontal systems, upper-level vortices, or jet streaks), favorable combinations of atmospheric instability and vertical wind shear can initiate tornadic storms. Moreover, complex topography and mesoscale boundaries, such as orographically-induced lifting and localized outflow interactions can enhance vertical motion and aid tornadogenesis \citep{Kirshbaum2018,Nixon2024}. In the tropics, recent studies have shed light on tornadic environments using Mexico as reference. \citet{LEONCRUZ2025} identified three types of tornadic settings, one of which is associated with the North American Monsoon (NAM)\footnote{In literature, they labeled it as Tornadic Environment Type 3 (TET3).}. This regime brings warm, moist inflow to the northwestern Mexico, resulting in modest-to-high instability ($\geqslant$ 1000 J kg$^{-1}$) and favorable kinematics (Deep Layer Shear/DLS $\sim$ 10-15 m s$^{-1}$; 0-3 km Storm-relative Helicity/SRH03 $\geqslant$ 100 m$^{2}$ s$^{-2}$). Mexico's tornado season, spanning April to September with peaks in May and July, bears a strong similarity to that of the Philippines. In both countries, monsoon activity, tropical cyclones, and easterly waves contribute to the generation of severe weather \citep{Leon-Cruz2022,Capuli2024a}.

On 27 May 2024, shortly after midday, a discrete supercell thunderstorm developed over the eastern section of Pampanga province. This storm subsequently produced a tornado that touched down in the suburb of Candating, located within the municipality of Arayat. Due to the rarity of supercell tornadoes in the Philippines,the Candating tornado was selected as a case study. This event offers a valuable opportunity to investigate the meteorological and dynamical processes associated with tornado initiation and development in a tropical setting. Although initial reports and media were scattered across various social platforms, Project SWAP facilitated the systematic collection, verification, and archiving of these materials. Eyewitness photographs and video footage, and damage reports were compiled to form a comprehensive dataset, presented in Section 3, which provides essential context for assessing the tornado’s impact. This study aims to examine the meteorological conditions that led to tornadogenesis, encompassing the synoptic, mesoscale, and dynamic environmental aspects, as well as the microphysical characteristics of the parent storm. To our knowledge, this is the first documented case of a tornadic supercell in the Philippines supported by extensive observational data and evaluated through sounding-derived parameters. As such, this investigation serves as a foundational study from which future research on tropical tornadoes can build. 

The paper is organized as follows; Section 2 outlines the materials and methods, including the reanalysis dataset, atmospheric profile parameters, satellite, radar, and lightning data used, in the study. Section 3 presents a comprehensive analysis of the visual-damage, environmental context (from synoptic-scale to mesoscale), climatological background, and storm dynamics associated with the event. Finally, Section 4 provides the summary of findings and an extended discussion of the results

\section{METHODS}

\subsection{Environmental Reanalysis Dataset}

To characterize the synoptic and mesoscale environment associated with the 27 May 2024 Pampanga tornado, this study utilized reanalysis data to assess the potential for SCS development. Reanalysis datasets are widely employed to diagnose atmospheric conditions conducive to severe weather such as tornadoes, damaging winds, and large hail. They are particularly useful for calculating instability indices and other parameters essential to SCS forecasting \citep{Taszarek2020}.

Environmental reanalysis data for this case were obtained from the fifth-generation European Centre for Medium-Range Weather Forecast (ECMWF) reanalysis (ERA5) accessed through the Climate Data Store \citep[CDS;][]{Hersbach2020}. ERA5 has demonstrated good performance in depicting vertical profiles of convective environments, especially across the United States and Europe \citep{Coffer2020,Taszarek2021a,Pilguj2022}. However, some known biases persist, particularly in the boundary layer. These include discrepancies in low-level parcel characteristics and vertical shear parameters, especially near surface boundaries \citep{King2019}. Such biases are influenced by geographic location and surface elevation. Despite these limitations, ERA5 remains among the most reliable and accessible datasets for investigating severe convective environments, globally \citep{Coffer2020,Taszarek2021b,Taszarek2021a}. 

For large-scale synoptic analysis, the study domain extended from 4°-22° N latitude and 110°-135° E longitude. The mesoscale sector, which also served for satellite visualization, covered 13°-19° N and 118°-125° E. Both domains used a spatial resolution of 0.25° x 0.25°. An hourly temporal resolution was used to allow detailed analysis of the atmospheric evolution surrounding the event, which occurred around 05 UTC. The time steps selected for in-depth analysis were 04, 05, and 06 UTC, enabling a focused examination of pre-, during, and post-storm conditions. 

\subsection{Vertical Profile: Skew-T Hodograph}

Observed proximity sounding within 200 km of the event were examined to characterize the convective and kinematic environment associated with the 27 May 2024 Tornado. An observed rawinsonde (RAOB) profile was obtained from the University of Wyoming (UWyo) database, specifically from the PAGASA Synoptic Station at Tanay, Rizal (WMO ID: 98433; 14.57° N, 121.37° E; elevation: 614 mASL) located approximately 86 km from the tornado initiation area. These observations are conducted routinely at 00 UTC and 12 UTC. In addition to the observed sounding, a model-derived profile was also extracted from ERA5 data using a combination of single-level and 137 hybrid sigma-pressure level fields. Both the observed and reanalysis-based soundings were analyzed to assess the thermodynamic and kinematic environment preceding the tornadic event.

The parcel profiles were computed assuming a non-entraining, irreversible adiabatic process, following the recent formulation by \citet{Peters2022}. Compared to the pseudoadiabatic ascent, which assumes all condensate is assumed to fall out of an air parcel immediately \citep{48}, the parcel profile now accounts for the layer of mixed-phase condensate in which liquid and ice are present just below the triple point temperature. This method is grounded in energy conservation, rather than conservation of moist entropy, reducing known biases inherent in other parcel models. In fact, \citet{Xu1989} showed that environmental temperature profiles in the tropics more closely align with an adiabatic parcel than pseudoadiabatic, making this approach relevant for analyzing deep convection in tropical settings.

Additionally, an entraining, irreversible adiabatic ascent was also employed to compute the Entraining CAPE \citep[ECAPE;][]{Peters2023b}. ECAPE accounts for dilution of updraft plumes due to environmental entrainment, particularly under conditions of limited storm-relative inflow and/or mid- to upper-tropospheric dryness. Thus, providing a more realistic estimate of updraft intensity than the conventional, undiluted CAPE, which can overestimate buoyancy in environments where entrainment is significant. All undiluted buoyancy profiles are calculated using virtual temperature correction. Depending on meteorological context, either most-unstable (MU) and surface-based (SB) parcels were selected for key thermodynamic parameters. CAPE and the level of free convection (LFC) were computed using MU parcel profiles, while convective inhibition (CIN) and lifted condensation level (LCL) were determined using SB profiles. Although mixed-layer parcels are often preferred for their robustness, SB and MU parcels were preferred in this study due to their better representation of near-surface instability (or stability) and lowest cloud base heights. These factors are critical in distinguishing surface-based from elevated convection, particularly in tornadic environments. 

The hodograph was used to understand the kinematic properties associated with the tornado event. Conventional hodographs apply a ground-relative approach where no transformation is made to the modeled $u$ and $v$ wind components \citep{Markowski2003,Nowotarski2018}. However, this method cannot capture how the wind profile changes from the perspective of the storm or assess for diagnosing convective storm dynamics. To address this, the study introduces the storm-relative hodograph (SR hodograph), a novel analytical approach within the Philippine meteorological context. This was done by subtracting the estimated storm motion calculated using the Bunkers ID Method \citep[B2K; also an important component of ECAPE;][]{Bunkers2000}, from the modeled wind profile. The resulting storm-relative winds (SRW, represented as V$_{SR}$) was then analyzed based on assumptions regarding storm motion. Due to the absence of observational constraints such as storm motions as derived from radar, supercell type, or storm mode, the right-moving vector was assumed, consistent with standard practice for the Northern Hemisphere. This assumes that the storm moves to the right of the non-pressure-weighted mean wind. 

The kinematic parameters were calculated from the interpolated model output of the u and v wind components. The bulk wind difference (BWD) was calculated across several vertical layers to assess shear magnitude. These included the layers from  0–500 m (BWD$_{500}$), 0–1 km (BWD$_{01}$), 0–3 km (BWD$_{03}$), 0–6 km (BWDD$_{06}$), 1–3 km (BWD$_{13}$), and 1–6 km (BWD$_{16}$). The vertically integrated storm-relative flux of streamwise vorticity into an updraft, represented by the storm-relative helicity (SRH), is of the greatest dynamical importance in this regard \citep{DaviesJones1984}. The SRH was computed for several layers: 0–500 m (SRH$_{500}$), 0–1 km (SRH$_{01}$), 0–3 km (SRH$_{03}$), and 1–3 km (SRH$_{13}$). For the near-ground layer (0–1 km), the mean V$_{SR}$ was calculated relative to the B2K$_{\text{RM}}$. This analysis also included the mean streamwise vorticity integrated over 0-500 m and 0-1 km, as well as the maximum streamwise vorticity within the lowest kilometer, denoted as \textgreek{ω}$_{\text{s500}}$, \textgreek{ω}$_{\text{s01}}$, and \textgreek{ω}$_{\text{smax}}$, respectively. These quantities were calculated as:

\[\omega_s = \frac{\nabla \times (v-c) \cdot (v-c)}{||(v-c)||} \]

where $v$ is the three-dimensional environmental wind and $c$ is the storm motion vector estimated through the B2K method. In addition, the vertical vorticity along the low-level mesocyclone was calculated. Following \citet{Peters2023a}, and based on the first-order approximation of tilting efficiency proposed \citet{Coffer2023}, the horizontal vorticity generated from the tilting of streamwise vorticity in a purely streamwise environment can be expressed as;

\[\zeta_{\text{LLM}} = \omega_s\frac{w}{V_{\text{SR}}} \]

\begin{table*}[h!t!]
\caption{Environmental Sounding Parameters to be evaluated in this case.}
    \centering
    \begin{tabular}{llll}
    \hline\hline
    Parameter & Definition & Notes & Units \\
    \hline
    CAPE & Convective Available Potential Energy & MU parcel & J kg$^{-1}$ \\
    CAPE$_{03}$ & 0-3 km CAPE & MU parcel & J kg$^{-1}$ \\
    CAPE$_{06}$ & 0-6 km CAPE & MU parcel & J kg$^{-1}$ \\
    ECAPE & Entraining CAPE & MU parcel & J kg$^{-1}$ \\
    CIN & Convective Inhibition & SB parcel & J kg$^{-1}$ \\
    LCL & Lifted Condensation Level & SB parcel & m \\
    LFC & Level of Free Convection & MU parcel & m \\
    LR$_{03}$ & 0-3 km Lapse Rate & Low-level Lapse Rate & $^{\circ}$C km$^{-1}$ \\
    LR$_{36}$ & 3-6 km Lapse Rate & Mid-level Lapse Rate & $^{\circ}$C km$^{-1}$ \\
    RH$_{13}$ & Mean 1-3 km Relative Humidity & & frac / $\%$\\
    RH$_{16}$ & Mean 1-6 km RH & & frac / $\%$\\
    BWD$_{01}$ & 0-1 km Bulk Wind Difference & Low-level Shear (LLS) & kt / m s$^{-1}$ \\
    BWD$_{03}$ & 0-3 km BWD & Mid-level Shear (MLS) & kt / m s$^{-1}$ \\
    BWD$_{06}$ & 0-6 km BWD & Deep-layer Shear (DLS) & kt / m s$^{-1}$ \\
    BWD$_{13}$ & 1-3 km BWD & & kt / m s$^{-1}$ \\
    BWD$_{16}$ & 1-6 km BWD & & kt / m s$^{-1}$ \\
    B2K$_{\text{RM}}$ & Bunkers RM Storm Motion & RM: Right Mover & kt / m s$^{-1}$ \\
    V$_{SR}$ & Mean 0-1 km Storm-relative Winds & Using B2K$_{\text{RM}}$ & kt / m s$^{-1}$ \\
    SRH$_{500}$ & 0-500 m Storm-relative Helicity & Using B2K$_{\text{RM}}$ & m$^{2}$ s$^{-2}$ \\
    SRH$_{01}$ & 0-1 km SRH & Using B2K$_{\text{RM}}$ & m$^{2}$ s$^{-2}$ \\
    SRH$_{03}$ & 0-3 km SRH & Using B2K$_{\text{RM}}$ & m$^{2}$ s$^{-2}$ \\
    SRH$_{13}$ & 1-3 km SRH & Using B2K$_{\text{RM}}$ & m$^{2}$ s$^{-2}$ \\
    $\omega_{s500}$ & 0-500 m Streamwise Vorticity & Using B2K$_{\text{RM}}$ & s$^{-1}$ \\
    $\widetilde{\omega_s}$$_{500}$ & 0-500 m Streamwiseness & Using B2K$_{\text{RM}}$ & frac / $\%$ \\
    $\omega_{s01}$ & 0-1 km Streamwise Vorticity & Using B2K$_{\text{RM}}$ & s$^{-1}$ \\
    $\widetilde{\omega_s}$$_{01}$ & 0-1 km Streamwiseness & Using B2K$_{\text{RM}}$ & frac / $\%$ \\
    $\omega_{\text{max}}$ & 0-1 km Maximum Streamwise Vorticity & Using B2K$_{\text{RM}}$ & s$^{-1}$ \\
    $\widetilde{\omega}$$_{\text{max}}$ & 0-1 km Maximum Streamwiseness & Using B2K$_{\text{RM}}$ & frac / $\%$ \\
    $\zeta_{\text{LLM}}$ & Low-level Vertical Vorticity (0-1 km) & MU parcel and B2K$_{\text{RM}}$ & s$^{-1}$ \\
    CA & Critical Angle & Using B2K$_{\text{RM}}$ & degree \\
    DTM & Deviant Tornado Motion & Using B2K$_{\text{RM}}$ & kt / m s$^{-1}$ \\
    WMAXSHEAR & Undiluted Updraft Velocity $\times$ BWD$_{06}$ & MU parcel & m$^{2}$ s$^{-2}$ \\
    \hline
    \end{tabular}
    \label{table:1}
\end{table*}

where \textgreek{ω}$_{\text{s}}$ is streamwise vorticity\footnote{In this case, angle-average and its maximum \textgreek{ω}$_{\text{s}}$ in the lowest 1 km.}, $w$ is the vertical velocity of the updraft\footnote{CAPE$_{03}$ will be the proxy for the vertical velocity in the low-level rotating updraft, assuming $w$ $=$ $\sqrt{2\text{CAPE}_{03}}$.}, and V$_{SR}$ is the storm-relative wind in the lowest 1 km. This formulation highlights how the conversion of environmental streamwise vorticity into vertical vorticity depends on the interaction and balance between vertical and horizontal motions \citep{DaviesJones1984}. It provides a physically intuitive approximation for assessing the potential for mesocyclone development and subsequent tornadogenesis.

Other kinematic parameters were also evaluated to characterize storm dynamics. The “critical angle” (CA), which describes the angle between the storm inflow and the shear vector in lowest kilometer, was calculated using the method proposed by \citet{Esterheld2008}. The “streamwiseness” (\textgreek{ῶ}$_{\text{s}}$) of horizontal vorticity was also calculated by dividing the integrated streamwise component of horizontal vorticity by the integrated total magnitude of the horizontal vorticity in a given layer;

\[\widetilde{\omega_s} = \frac{\omega_s}{\omega} \]

This formulation is conceptually similar to the streamwise-to-crosswise ratio used in tornado environment studies such as \citet{Coffer2019}. However, a key distinction is that while the streamwise-to-crosswise vorticity ratio approaches infinity when the crosswise components approaches zero, which is a relatively common condition in tornado environments, the \textgreek{ῶ}$_{\text{s}}$ parameter approaches infinity only when the horizontal shear within the layer approaches zero. Although neither behavior is ideal, the use of \textgreek{ῶ}$_{\text{s}}$ is justified for two reasons. First, near-zero shear conditions are rare in tornado environments \citep{Coffer2020}. Second, the parameter offers intuitive interpretation; for example, a value of 0.9 indicates 90\% of the horizontal vorticity is streamwise in nature. To the author’s knowledge, this unitless parameter has not yet been applied in statistical analyses of supercells. Nonetheless, the degree to which vorticity is streamwise remains a fundamental aspect of supercell dynamics and tornadogenesis \citep{Coffer2017}.

Lastly, the Deviant Tornado Motion (DTM) was calculated based on the method developed by \citet{Nixon2021}. This parameter evaluates the potential deviation of tornado motion from both the low-level advective flow and the parent storm motion, and offers insight into tornado track behavior relative to ambient wind conditions. All soundings associated in this event were analyzed using SounderPy (v3.0.8) by \citet{Gillett2025}. A list of the evaluated kinematic and thermodynamic parameters, along with their definitions and computation methods, is presented in Table \ref{table:1}.

\subsection{Spaceborne Data}
\subsubsection{HIMAWARI-9 Satellite Data}

\begin{figure}[t]
\centering
\includegraphics[width=\columnwidth]{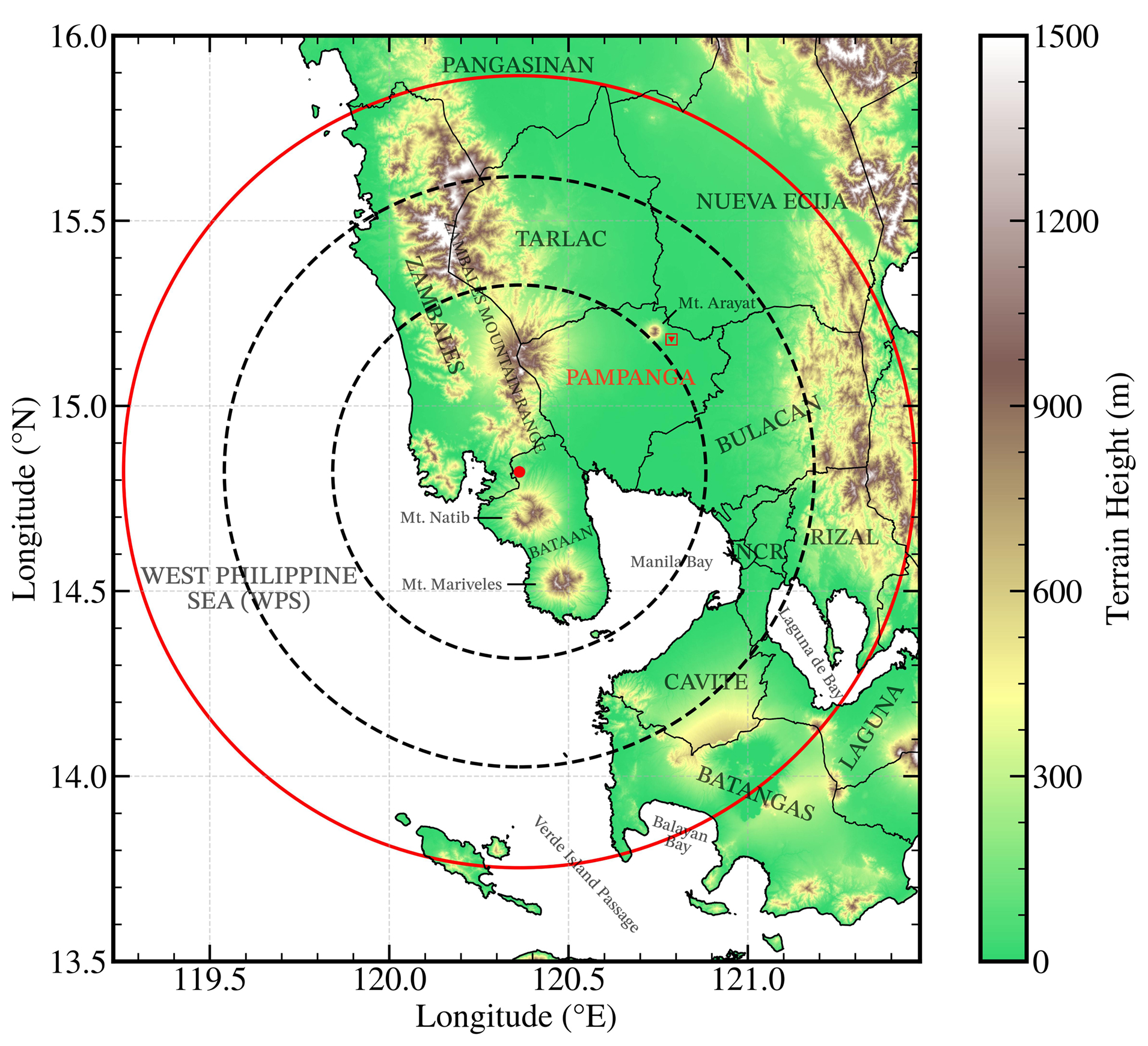}
\caption{Elevation map of Metro Manila (NCR) and surrounding provinces. S-SUB radar is shown as a small red circle, with 120 km radar extent coverage depicted as a red circle (solid line).  The red triangle enclosed by a box is the location of the tornadic event}\label{fig1}
\end{figure}

High spatio-temporal resolution data from the Advance Himawari Imager (AHI) aboard the HIMAWARI-9 satellite \citep{Bessho2016} were used to identify and characterize the convective system associated with the tornadic event. HIMAWARI-9 comprises of 3 VIS bands (central $\lambda$ ranging from 0.47 {\textmu}m$-$0.64 {\textmu}m), 3 NIR bands (central $\lambda$ ranging from 0.86 {\textmu}m$-$2.3 {\textmu}m), and 10 IR bands (central $\lambda$ ranging from 3.9 {\textmu}m$-$13.3 {\textmu}m). Particularly, this study selected the visible bands and the 10.4 {\textmu}m  infrared) atmospheric window band (Band 13; B13) to monitor deep convection. B13 is particularly useful for detecting cold cloud-top regions [Brightness Temperature (BT) $<$ 235 K ($-$40 °C)] while minimizing interference from  water vapor \citep{Hirose2019}. It is also widely adopted for rapid identification of convective features associated with extreme rain events with small lag time \citep{Wang2022}. The visible bands have  spatial resolution ranging from  500 m$-$1 km, while B13 has a coarser resolution of 2 km.

To enhance visual interpretation of convective features, the Japan Meteorological Agency’s Daytime Convective RGB composite \citep[JMA-RSMC Tokyo;][]{Yamashita_2018} was also employed. This RGB product combines spectral differences that highlight cloud phase and vertical development. The Red component is based on the B10-B08 difference, Green on B13-B07, and Blue on B03-B05. These combinations help identify presence of ice in clouds and the potential for hail-producing updrafts. All satellite-derived datasets used in the study were based on 10-minute full-disk scans from HIMAWARI-9. 

Penetrative overshooting tops (OT) were also examined as signatures of severe convection. OTs are identified as local minima in cloud-top brightness temperature (BT $<$ $-$60 °C) and signify strong updrafts capable of reaching above the equilibrium level (EL) and into the lower stratosphere. Their occurrence is frequently linked to intense strom dynamics and severe weather, including large hail and tornadoes \citep{Bedka2010}.

\subsubsection{Landsat-9 OLI}

Landsat-9 Operational Land Imager (OLI) data with 30 m spatial resolution was acquired from the U.S. Geological Survey (USGS) Earth-Explorer platform to delineate the tornado damage track. Pre- and post- event images were obtained on 01 May and 02 June 2024, respectively. To ensure consistency across the two data collection dates, radiometric correction was performed by converting the raw multispectral brightness digital number (DN) values to top of atmosphere (TOA) reflectance. This normalisation procedure is crucial for multi-temporal analysis, as it compensates for sensor calibration differences, Earth-Sun distance variability, and changes in solar zenith angle. While the DN-to-TOA conversion accounts for sensor-specific gains and solar geometry, it does not correct for atmospheric scattering effects. To address atmospheric distortion, the Dark Object Subtraction (DOS) method was applied \citep{Chavez1988,SONG2001}. This method estimates and removes the scattered path radiance component, thereby enhancing the reliability of surface reflectance interpretation. To detect changes in vegetation and surface condition due to the tornado, the well established normalized difference vegetation index (NDVI) was used \citep{TUCKER1979,Singh1989}. NDVI is defined as:

\[\text{NDVI} = \frac{\text{NIR}-\text{red}}{\text{NIR}+\text{red}} = \frac{\text{Band 5}-\text{Band 4}}{\text{Band 5}+\text{Band 4}} \]

To improve spatial clarity, all associated bands were pan-sharpenned using Band 8 (also known as Panchromatic Band) through the Simple Mean method, a basic component substitution technique. The resulting pan-sharpened images were saved in 16-bit unsigned integer TIFF (.tif) format for further geospatial analysis. This approach aligns with prior studies on tornado and storm damage assessment using Landsat and other optical remote sensing techniques \citep{Wilkinson2010,Gallo2012,Wagner2012,Molthan2014}. 

\begin{table}[h!t!]
\caption{Technical specifications of Subic Radar (S-SUB).}
\centering
\resizebox{\columnwidth}{!}{%
    \begin{tabular}{ll}
    \hline\hline
    Bandwidth & S-band \\
    Polarization & Single-Polarization (single-pol)\\
    Position & 14.822$^{\circ}$N, 120.363$^{\circ}$E \\
    Altitude & 532 masl \\
    Pulse Width/Maximum Range & 0.8 ms/120 km \\
    Beam width/Range resolution & 1$^{\circ}$/500 m  \\
    Number of Elevation Angles & 14 \\
    Elevation Angles & 0.5$^{\circ}$, 1.5$^{\circ}$, 2.4$^{\circ}$, 3.4$^{\circ}$, 4.3$^{\circ}$, 5.3$^{\circ}$, 6.2$^{\circ}$, \\
    & 7.5$^{\circ}$, 8.7$^{\circ}$, 10$^{\circ}$, 12$^{\circ}$, 14$^{\circ}$, 16.7$^{\circ}$, 19.5$^{\circ}$ \\
    Volume Cycle Interval & 10 minutes \\
    Start of operation & 2012 \\
    \hline
    \end{tabular}%
}
\label{table:2}
\end{table}

\subsection{Observational Data from DOST-PAGASA}

\subsubsection{Radar Data}

DOST-PAGASA operates a nationwide network of 17 weather radars, 7 of which are dual-polarization (dual-pol) C-band radars, 3 dual-pol S-band, and 7 single-polarization (single-pol) S-band radars. These systems support the agency’s capacity for early detection and monitoring of weather systems across the archipelago.  For this study, data were obtained from the Subic S-band radar (S-SUB), a single-polarization radar strategically located on an elevated site (532 masl) in Subic, Bataan (14.822° N, 120.363° E); northwest of Metro Manila (Fig. \ref{fig1}). The S-SUB radar provides extensive coverage over Metro Manila and the surrounding Central Luzon region, an area that receives high mean annual rainfall. S-SUB employs a Modified Volume Coverage Pattern 11 (VCP11) which optimizes radar sampling across both low and high elevation angles. This configuration is particularly well-suited to tropical environments. For this study, the Doppler mode scanning radius was limited to 120 km to balance the trade-off between the maximum Nyquist velocity and range of spatial coverage. Technical radar specifications are described in Table \ref{table:2}.

Radar data underwent quality control based on a fuzzy logic approach designed to detect and eliminate interference patterns. This method draws on prior algorithm assessments \citep[e.g.,][]{Berenguer2006,YoHan2006,Gourley2007,Kilambi2018,Overeem2020} and was adapted for both single and dual-pol radar products following the framework described in \citet{Lin2021}. Velocity de-aliasing was also necessary due to the radar’s low unambiguous velocity threshold ($\sim$8.03 m s$^{-1}$), which is unusual for S-band radar systems. The accuracy and confidence level of the radial velocity analysis in this study are inherently limited by the radar’s low Nyquist velocity. Since the Nyquist velocity defines the maximum unambiguous radial velocity measurable without aliasing, a low threshold (just like S-SUB has that time or even back then) makes the S-SUB data particularly susceptible to velocity folding in regions of strong wind shear and/or intense rotation is/are present.

\begin{table}[h!t!]
\caption{Mesocyclone detection thresholds following \citet{Stumpf1998}, \citet{Hengstebeck2018}, and \citet{Feldmann2021}.}
\centering
\resizebox{\columnwidth}{!}{%
    \begin{tabular}{lll}
    \hline\hline
    Variable & Minimum Threshold & Maximum Threshold \\
    \hline\hline
    Rotational Velocity & 6 m s$^{-1}$ & 10 m s$^{-1}$\\
    Vorticity & $6 \times 10^{-3}$ & $6 \times 10^{-2}$ \\
    Duration & 10-12 minutes & N/A \\
    Horizontal Scale & 1 km & 10 km \\
    \hline
    \end{tabular}%
}
\label{table:3x}
\end{table}

To correct for velocity folding, a regional de-aliasing technique was implemented using the Python ARM Radar Toolkit \citep[PyArt;][]{Helmus2016}. The \textit{dealias region based} module was applied to identify coherent velocity regions, which were iteratively unfolded via dynamic network reduction methods based on local standard deviations. Specifically, we perform a two-stage de-aliasing and composite reconstruction of the radial velocity by identifying tornadic and non-tornadic signatures. A velocity texture field was computed using the \textit{calculate velocity texture} function, representing the spatial standard deviation of radial velocity. Regions with low velocity texture values ($<$ 5 m s$^{-1}$) were masked, and an initial de-aliasing of the entire velocity field was performed to obtain a smooth, coherent non-tornadic velocity background. The tornadic signature was then isolated using a simple gate filter that excluded reflectivity values below 40 dBZ. A second de-aliasing process was subsequently applied to these isolated regions, followed by an additional correction step using the \textit{dealias unwrap phase} module to remove residual artifacts introduced by the region-based de-aliasing. Finally, the non-tornadic and tornadic de-aliased velocity fields were merged into a composite de-aliased velocity product, achieved by masking invalid or missing values in the non-tornadic field and replacing them with corresponding values from the tornadic field. While the applied two-stage de-aliasing quality control procedure for the radial velocity mitigates many of these issues, the low Nyquist velocity fundamentally constrains the fidelity of the retrieved radial velocity field. Consequently, rotational velocity estimates derived from the S-SUB should be interpreted with caution, particularly when assessing magnitudes or fine-scale circulation features.

To identify potential rotation signatures within the de-aliased radial velocity field of the S-SUB radar, we used the Mesocyclone Detection Algorithm (MDA) and its detection thresholds as developed by \citet{Stumpf1998} and following modern applications of this MDA both as in \citet{Hengstebeck2018} and \citet{Feldmann2021}. These parameters involved the maximum tangential (rotational) velocity, vorticity, horizontal scale, and duration of the feature. In particular, the rotational velocity (per nomenclature; \text{V$_{\text{rot}}$}) is calculated using the maximum outbound velocity (\text{V$_{\text{max,out}}$}) and the maximum inbound velocity (\text{V$_{\text{max,in}}$}) determined from the segments in the 2D feature: 

\[
\text{V$_{\text{rot}}$} = \frac{\text{V$_{\text{max,out}}$}-\text{V$_{\text{max,in}}$}}{2}
\]

The vertical vorticity, as measured through the radar ($\zeta_{\text{rad}}$), can be determined using the equation below: 

\[
\text{$\zeta_{\text{rad}}$} = 2\frac{\text{V$_{\text{max,out}}$}-\text{V$_{\text{max,in}}$}}{\Delta\text{X}}
\]

where $\Delta$\text{X} is the cartesian distance (m) between the velocity maxima. 

While we apply these mesocyclone detection techniques and thresholds, we must assert that these variables are largely adapted from studies conducted in mid-latitude regions, where radar configurations, storm structures, and environmental conditions differ substantially from those in the tropics. Therefore, we encourage future readers (and more importantly, Filipino meteorologists who wanted to dive in severe weather meteorology) to develop more robust and regionally calibrated detection algorithms suited to the Subic radar (S-SUB) and the broader Philippine radar network such as conducted by \citet{Feldmann2021} for the Alpine region. The thresholds for these radial velocity-derived parameters are summarized in Table \ref{table:3x}.

\begin{figure*}[t]
\centering
\includegraphics[width=\textwidth]{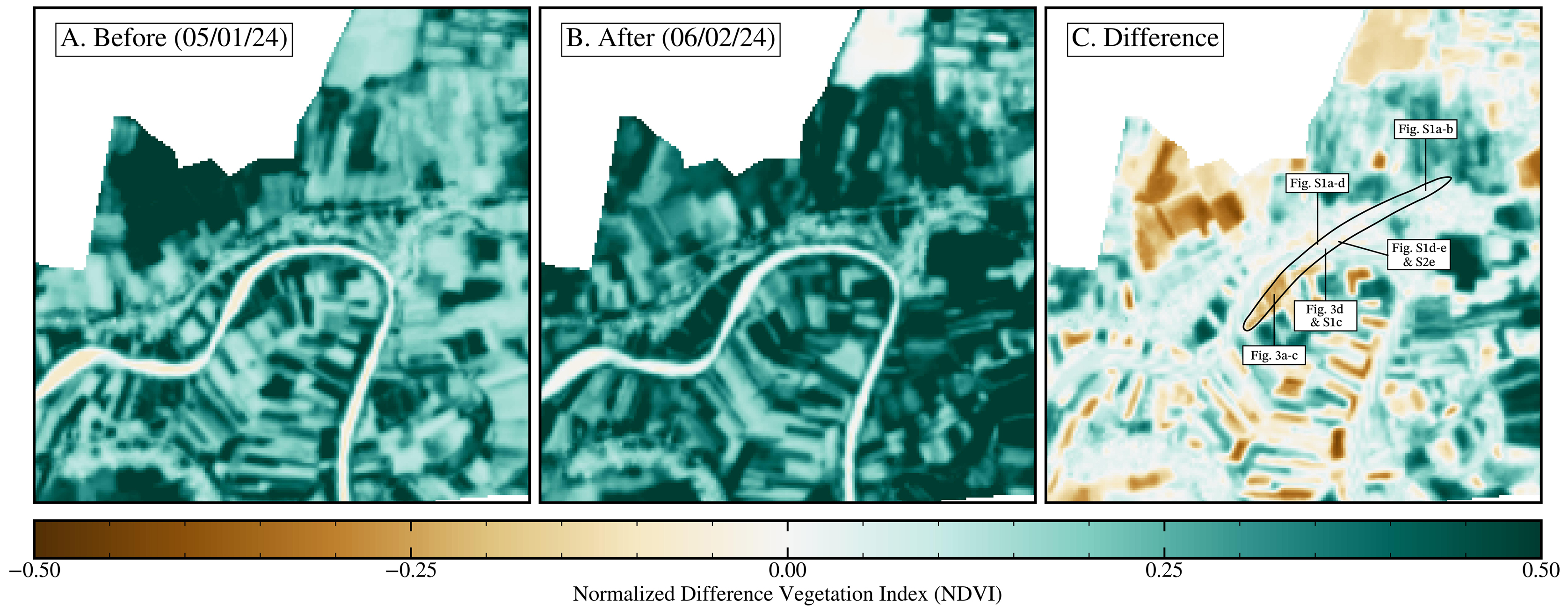}
\caption{Landsat-9 OLI Damage Extent of the Arayat, Candating Tornado. (a) Before the event at 01 May 2024, (b) After the event at 02 June 2024, and (c) Difference between the NDVI before and after. Highlighted is the potential damage track of the tornado}
\label{fig2}
\end{figure*}

The quality controlled radar data were interpolated onto a Cartesian grid with 500 m horizontal resolution using PyART. Grid points are found by interpolation of all radar gates using a double-pass Barnes weighting function within a constant radius of influence \citep[ROI;][]{Barnes1964,Pauley1990}. From the gridded Plan Position Indicator (PPI) reflectivity ($Z$) fields, hydrometeors were classified into three bulk ice-phase categories: graupel, hail, and ice (the latter referring specifically to ice crystals and vertically aligned ice). The graupel and hail categories include both low- and high-density graupel as well as hailstones. To estimate hydrometeor mass concentrations, reflectivity-mass ($Z‐M$) relationships were applied derived by \citet{Heymsfield1986} and \citet{Heymsfield1988}, and subsequently used in severe weather studies \citep{Deierling2008,Rocque2024}. The equations are as follow:

\[
\text{M$_{\text{ICE}}$} = 0.017 \times \textit{Z}^{0.529}
\]

\[
\text{M$_{\text{GRAUPEL}}$} = 0.0052 \times \textit{Z}^{0.5}
\]

\[
\text{M$_{\text{HAIL}}$} = 0.000044 \times \textit{Z}^{0.71}
\]

where $M$ is hydrometeor mass in g m$^{‐3}$ and $Z$ is the reflectivity in linear units (mm$^6$ m$^{‐3}$). While region-specific $Z‐M$ relationships have been developed in the U.S. and Japan, no such relations have been developed for the tropical environment of the Philippines. \citet{Deierling2008} compared several different $Z‐M$ relationships and demonstrated that while  absolute mass estimates may vary by an order of magnitude depending on the chosen $Z‐M$ relationship, the relative trends remain consistent. As such, the $Z‐M$ relationships from \citet{Deierling2008} were adopted for this study.

\subsubsection{Lightning Data}

Lightning data was obtained from DOST-PAGASA Lightning Detection Network (PLDN), which consists of 28 strategically deployed lightning sensors distributed throughout the country. These sensors are primarily located within PAGASA observing stations  and support real-time thunderstorm monitoring.  Since its operational inception in 2018, the PLDN has provided continuous coverage for both intra-cloud (IC) and cloud-to-ground (CG) lightning discharges. It also enables early detection of rapid increases in lightning activity, known as ‘lightning jumps’, which often precede the onset of severe weather \citep{Schultz2011}. For this case study, lightning data spanning from 03 to 07 UTC on 27 May 2024 were used. 

To ensure data quality, a spatial filter was applied to isolate lightning activity directly associated with the tornadic supercell. A geographic bounding box was defined from 120.75° E$-$121.10° E longitude and 15.00° N$-$15.20° N latitude, centered on the storm region. This filter excluded lightning unrelated to the storm and helped to eliminate signals from background convection. Within this defined domain, all lightning strokes were aggregated into 1-minute intervals. For each interval, we calculated the total number of strokes, along with average and median peak current values, separately for positive and negative polarities. This time-resolved approach enabled a detailed characterization of lightning intensity, polarity distribution, and temporal  evolution in relation to the supercell’s development and maturity. 

\begin{figure*}[t]
\centering
\includegraphics[width=\textwidth]{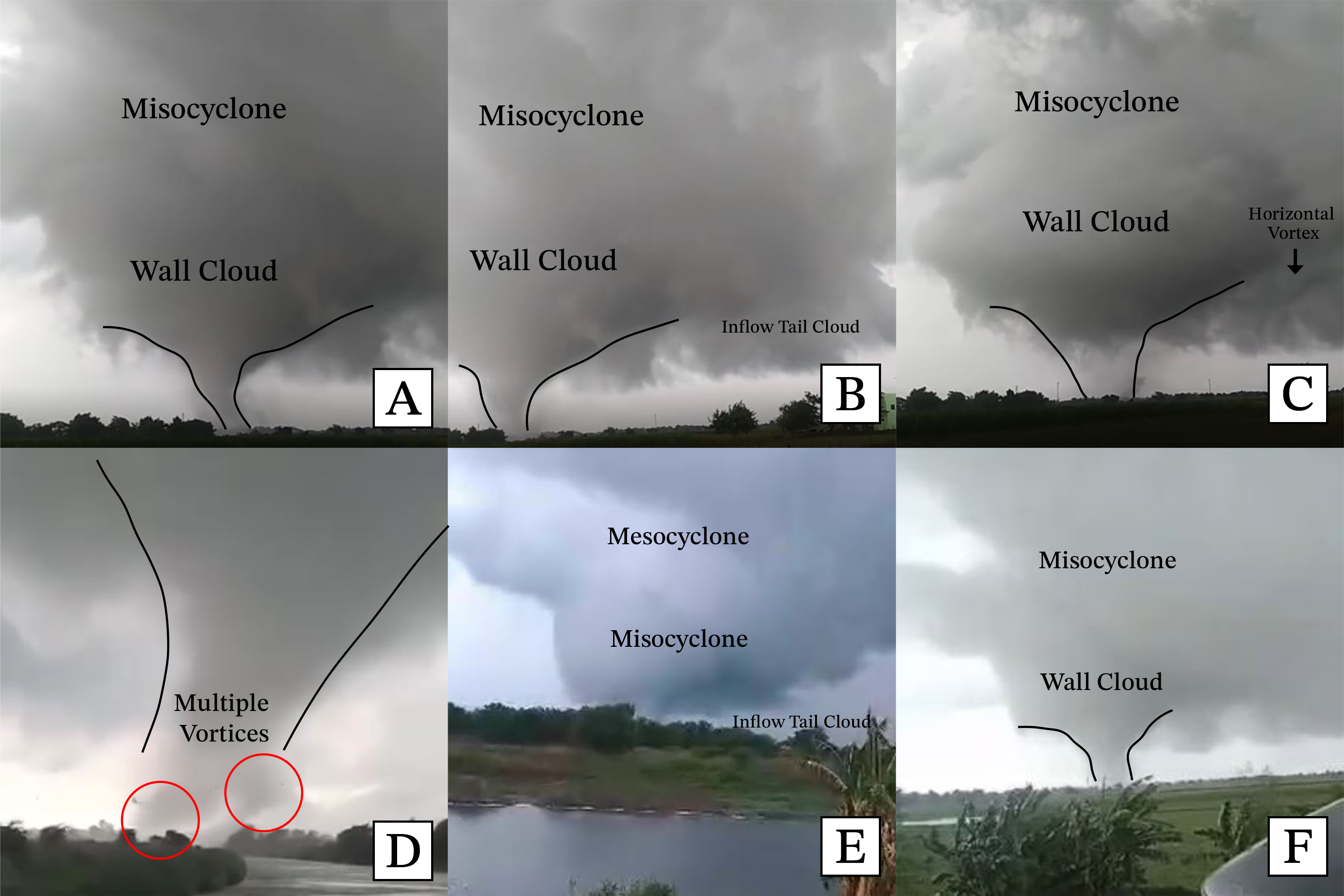}
\caption{Photographs of the 27 May 2024 Tornado that impacted Brgy. Candating in Arayat, Pampanga. Courtesy of (a-c) Joel Ulanday Soliman, (d) Aries Cruz, (e) Mark Guinto Estopin, and (f) Renz Reyes}
\label{fig3}
\end{figure*}

\section{RESULTS}

\subsection{Tornado damage path and visual observations}

The damage path of the tornado that struck Candating, Arayat was identified through change detection analysis using pre- and post-event Landsat-9 imagery. Figure \ref{fig2} illustrates the NDVI values before and after the event, as well as their difference. The study area mainly consists of land cover types of rural settings, including bare soil, residential structures, paved surfaces, and vegetation. A distinct southwest-to-northeast elongated zone, delineated by a black contour in Figures \ref{fig2}b and \ref{fig2}c exhibits the most significant change in NDVI between two image acquisition dates. This zone corresponds to the tornado’s damage path.  The variance in NDVI prior to the event was calculated at 0.5429 (dimensionless), which decreased to 0.2085 following the tornado. This represents an approximate 40\% reduction in NDVI variance, indicating a shift from a heterogeneous to a more homogeneous surface condition. Such a reduction is interpreted as a drying or scouring signal, consistent with widespread vegetation removal or ground exposure due to wind damage. Based on this analysis, the tornado carved a damage track approximately 2 km in length. The path originated in rice/crop fields and extended toward the populated center of Candating. Along this track, the tornado caused damage to various structures, including one- to two-story residential homes, multiple electrical transmission lines, a church, and a junior/senior high school building. These damages are consistent with the EF Scale criteria for EF2 intensity, as evaluated through corresponding Damage Indicators (DI) and Degrees of Damage (DoD) observed at multiple locations.  

\begin{figure*}[t]
\centering
\includegraphics[width=\textwidth]{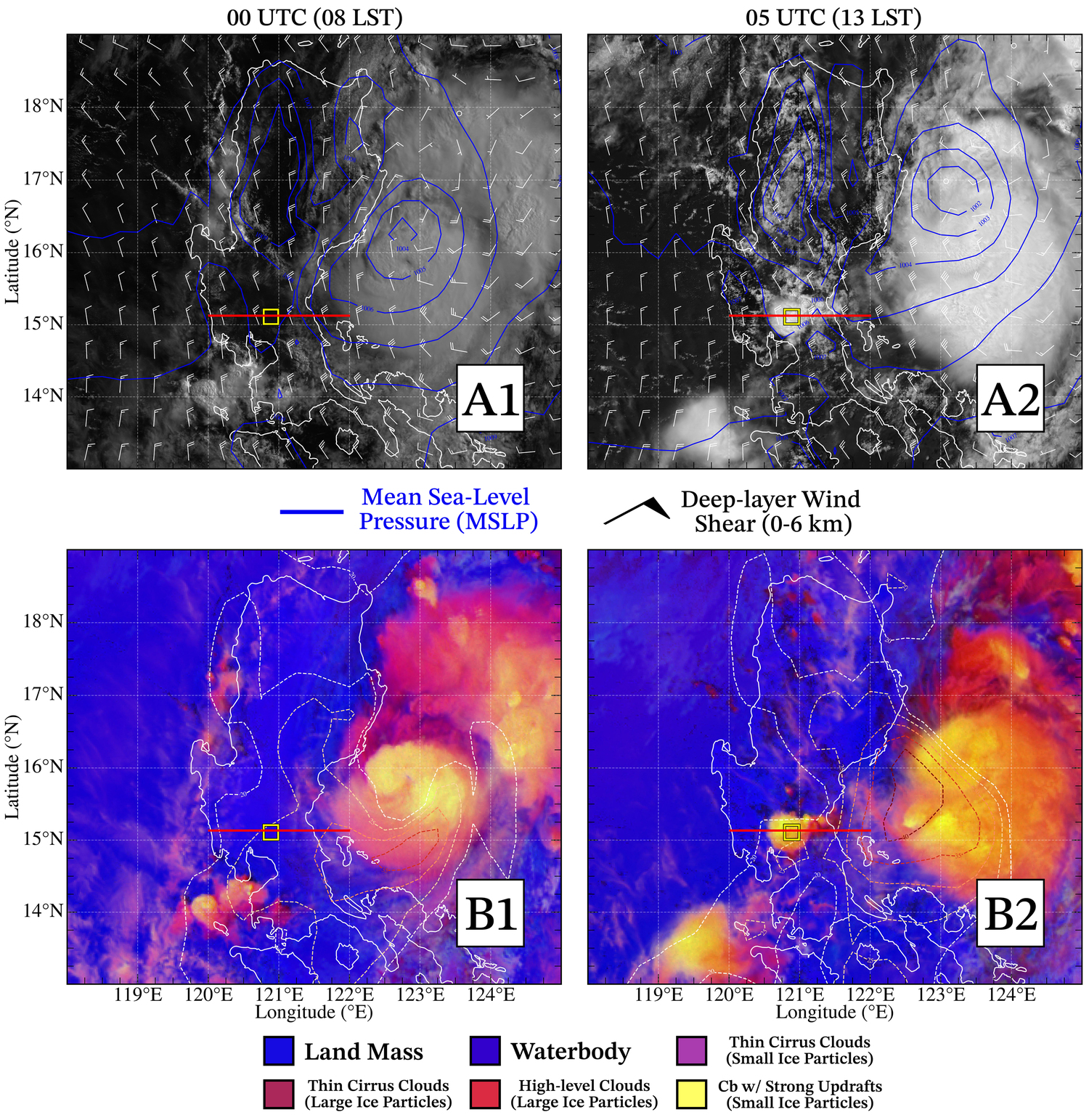}
\caption{HIMAWARI-9 AHI scans at (1) 00 UTC/08 LST and (2) 05 UTC/13 LST; (a) Band 03 0.64 μm with 0-6 km Shear Vectors (kt) and Mean Sea-Level Pressure (blue solid; hPa), and (b) Daytime Convective RGB with 0-6 km Shear Magnitudes $>$ 20 kts (dashed lines). The area of interest is demarcated in yellow box}
\label{fig4}
\end{figure*}

\begin{figure*}[t]
\centering
\includegraphics[width=\textwidth]{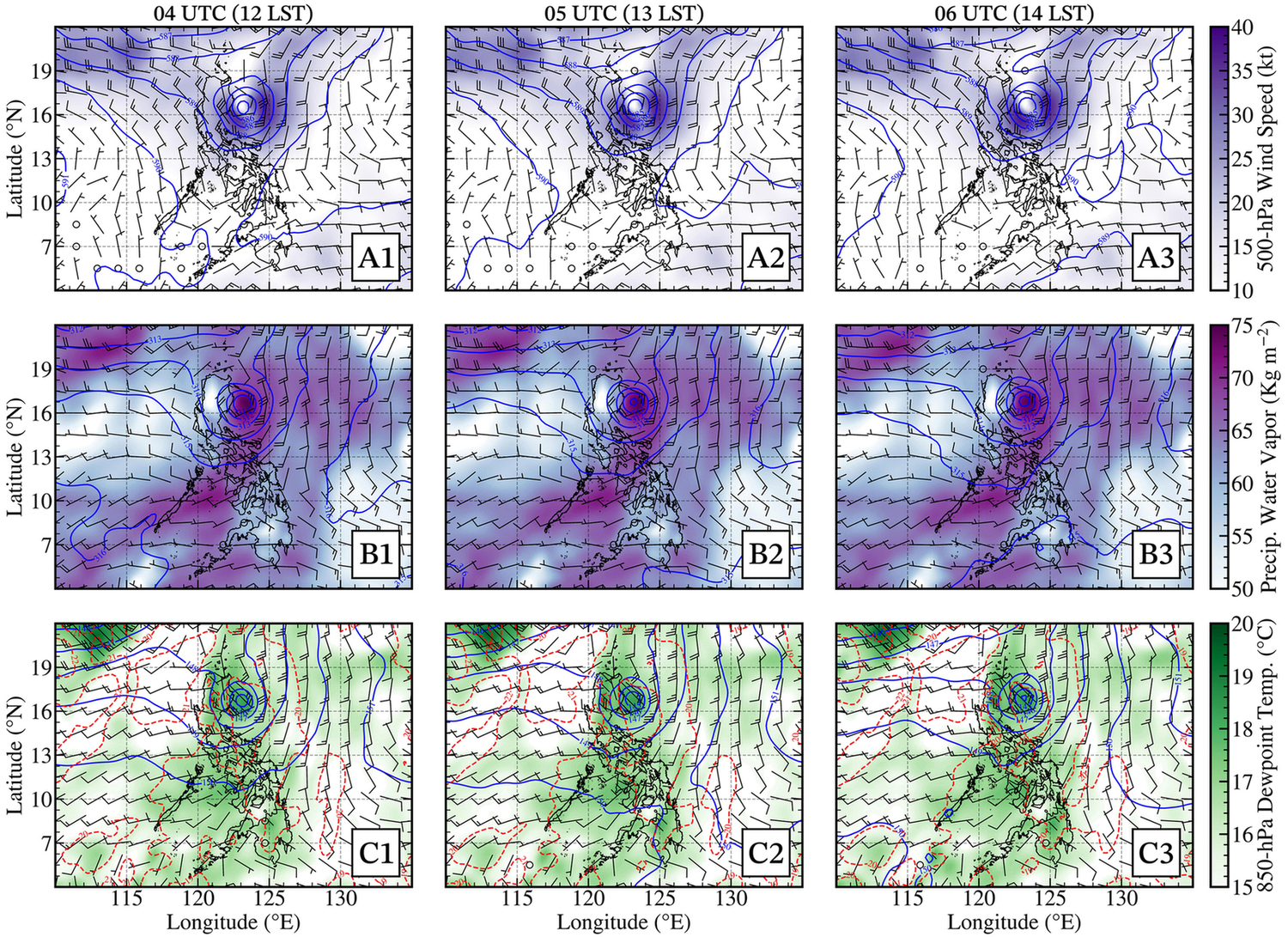}
\caption{Synoptic environment of the Philippine archipelago before (1; 04 UTC/12 LST), during (2; 05 UTC/13 LST), and after (3; 06 UTC/14 LST) the event. (a) 500-hPa Geopotential Heights (blue solid; dam) and winds (kt). Contours are wind speeds $>$ 10 kts. (b) 700-hPa Geopotential Heights (blue solid; dam), Precipitable Water Vapor $>$ 50 kg m$^{-2}$ (contour), and winds (kt). (c) 850-hPa Geopotential Heights (blue solid; dam), Dewpoints $>$ 15 $^{\circ}$C (contour), Temperature (red dashed; $^{\circ}$C), and winds (kt)}
\label{fig5}
\end{figure*}

In support of the EF2 classification, structural damage within the tornado-affected area was carefully analyzed using photographic evidence and the EF scale framework. One of the most compelling indicators was the complete collapse and removal of the curved steel roof of a recreational canopy located on school grounds (Fig. \ref{figS1}a). This structure was assessed under Damage Indicator (DI) 16 – \textit{Junior or Senior High School}, with a DoD of 8, corresponding to the uplift or collapse of a light steel roof structure. Based on the EF scale, this damage is consistent with estimated wind speeds of approximately 125 mph (56 m s$^{-1}$). In the same vicinity, a nearby school building sustained significant damage, with over 20\% of its metal roof decking uplifted (Fig. \ref{figS1}b). This was categorized as DoD 5 under the same DI, with an associated wind speed estimate of around 119 mph (53 m s$^{-1}$). Additional damage was observed near the church grounds, where multiple concrete and wooden electrical poles were found to be leaning or completely broken  (Fig. \ref{figS2}a-2c). These were categorized under DI 24 – \textit{Electrical Transmission Line}, with DoDs ranging from 4 to 5, indicative of wind speeds between 110 and 118 mph (49-53 m s$^{-1}$). Elsewhere along the path, several single-family residential structures experienced EF1 damage. In these cases, large portions, if not the entirety, of the roofing material were torn off (Figures \ref{figS1}d-e and S2e). This level of damage aligns with DoD 6 under DI 2 – \textit{One-or two-family residences}. Supplementary photographs of the affected structures are presented in Figures \ref{figS1} and \ref{figS2}, and are georeferenced across the NDVI-derived damage path shown in Figure \ref{fig2}. Notably, the spatial distribution of observed damages corresponds closely with in situ photographs provided by local residents. This alignment reinforces the reliability of the damage path delineation and further substantiates the EF2 tornado classification for this event. 

While the quality of the available photographs and videos limits the extent of detailed analysis, several visual indicators suggest that the parent storm exhibited supercellular characteristics. Most notably, the presence of a wall cloud and a possible misocyclone (Fig. \ref{fig3}a,b,f) indicates a broader low-level circulation near the storm base, which is consistent with the structure of a classic supercell \citep{Lemon1979}. The mesocyclone appears to reside above this misocyclone (Fig. \ref{fig3}e), although some associated features, such as rotating rain curtains typically linked to a rear-flank downdraft \citep[RFD;][]{Lemon1979,Markowski2002}, were not clearly observed. One particularly notable observation is a potential horizontal vortex orbiting the outer edge of the tornado at its peak intensity (Fig. \ref{fig3}c). While rarely documented in the Philippines, such features have occasionally been observed in strong tornadoes elsewhere and are believed to be manifestations of frictional or baroclinic processes within the storm outflow  \citep{Houser2016,Oliveira2019}. These mechanisms are known to play important roles in the dynamics of supercell tornadoes \citep{Fischer2020}. The tornado may have also exhibited a multiple-vortex structure, with subvortices potentially visible within the funnel (Fig. \ref{fig3}d). This feature is commonly associated with violent tornadoes \citep{Wurman2014}\footnote{A recent example of this is the Greenfield, Iowa Tornado (21 May 2024).}. While these interpretations are based on limited and partially obscured visual documentation, they offer important insights regarding the complexity of the vorticity-generating processes involved in the Candating, Arayat tornado. An additional annotated photograph highlighting key storm features is provided as supplementary in Fig. \ref{figS3}.

\begin{figure*}[t]
\centering
\includegraphics[width=\textwidth]{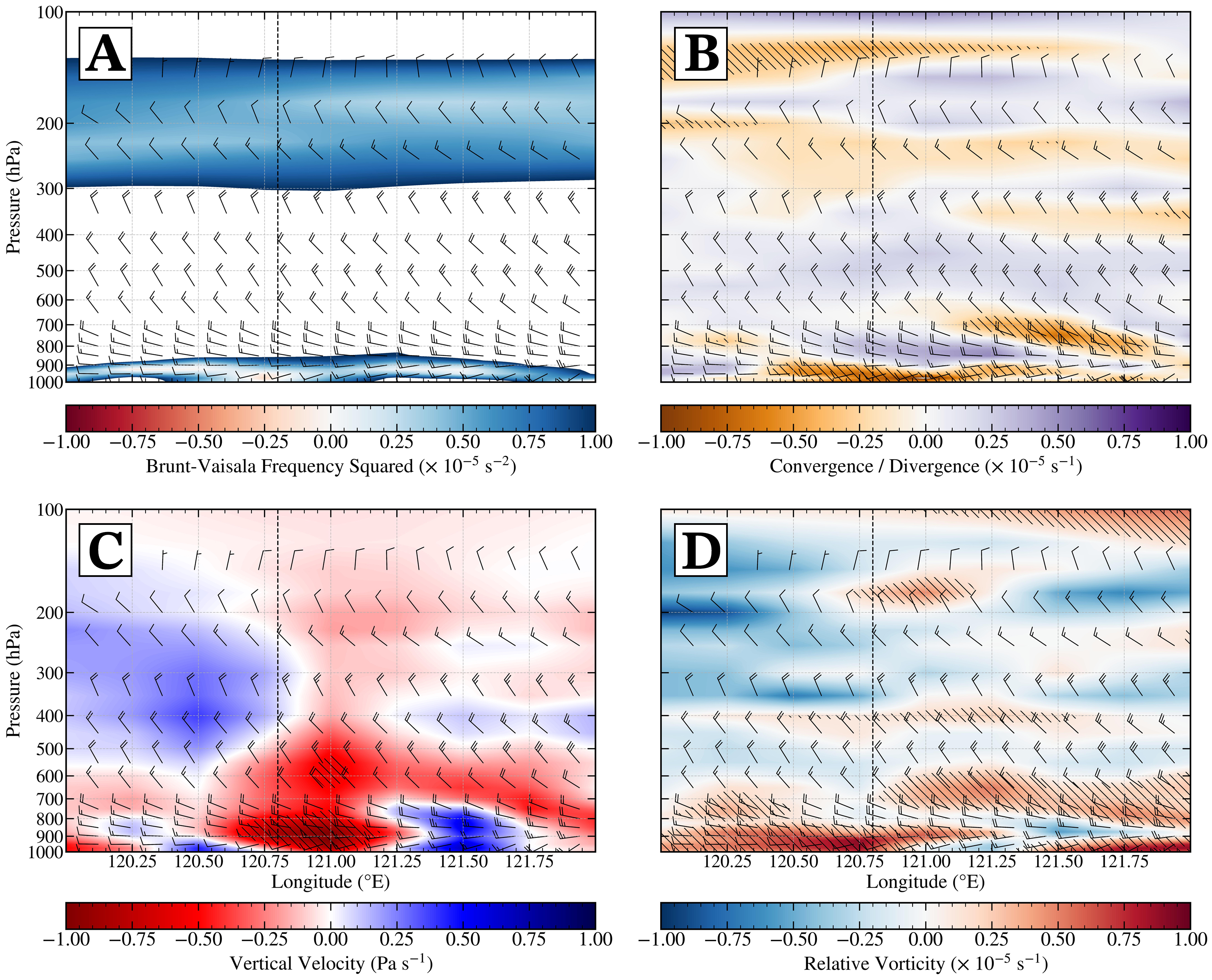}
\caption{05 UTC/13 LST ERA5 Vertical Cross Section of; (a) Brunt-Vaisala Frequency Squared ($\times$ 10$^{-5}$ s$^{-2}$), (b) Convergence / Divergence (Pa s$^{-1}$) with hatched area of Convergence $<$ $-$0.50 $\times$ 10$^{-5}$ s$^{-1}$, (c) Vertical Velocity (Pa s$^{-1}$) with hatched area of Convergence $<$ $-$0.50 Pa s$^{-1}$, and (d) Relative Vorticity ($\times$ 10$^{-5}$ s$^{-1}$) with hatched area of Absolute Vorticity $>$ 0.50 $\times$ 10$^{-5}$ s$^{-1}$. All cross sections are accompanied by tangential and normal components (kt)}
\label{fig6}
\end{figure*}

\subsection{Synoptic Environment}

In the morning hours of 27 May 00 UTC (08 LST), a key synoptic-scale feature was Tropical Cyclone Ewiniar (Assigned local name: TC Aghon), which has reached Typhoon intensity with a  central pressure of 965-hPa and maximum sustained winds of  140 km h$^{-1}$ (10-minute average). At this time, the cyclone was located in close proximity to mainland Luzon, particularly close to the province of Aurora, and was moving north-northeastward at approximately 10 km h$^{-1}$, gradually moving away from the landmass where it had previously made landfall.

\begin{figure*}[t]
\centering
\includegraphics[width=\textwidth]{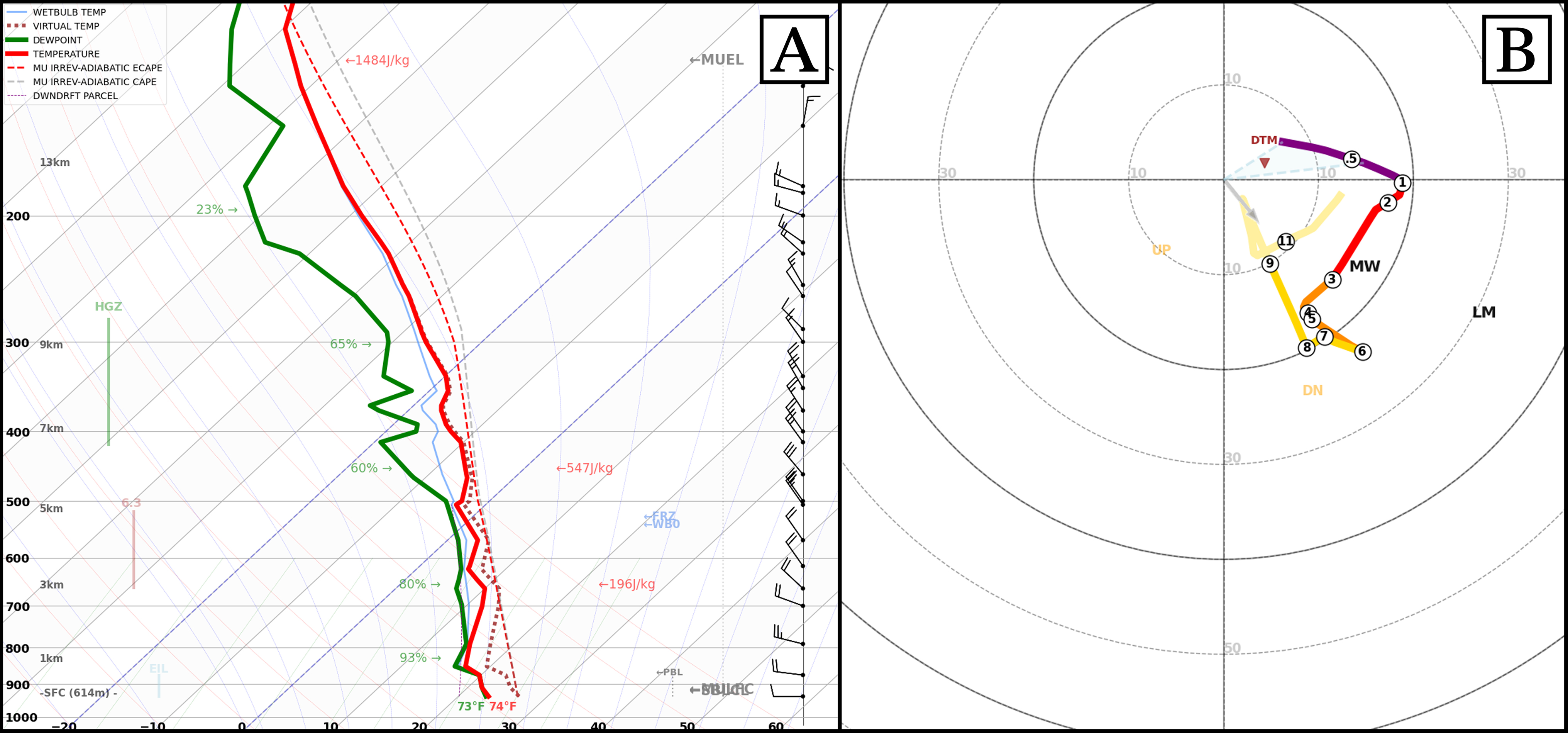}
\caption{00 UTC / 08 LST Tanay, Rizal observed proximity sounding composed of: (a) Thermodynamic profile and (b) associated Storm-relative Hodograph. Annotations to the thermodynamic profiles include the SBLCL, PBL, MULFC, FRZ, and MUEL. Meanwhile, annotations to the wind profile includes the individual Bunker’s Storm Motion, Deviant Tornado Motion, and Corfidi’s MCS components}
\label{fig7}
\end{figure*}

As shown in Figures \ref{fig4}a1 and \ref{fig4}b1, another important synoptic factor that aided in the severe convective initiation was the relatively cloud-free environment over the area of interest. This setup resembles the moist, open warm sectors commonly observed ahead of frontal systems and is often conducive to severe storm initiation. The lack of cloud cover allowed for sustained diurnal surface heating throughout the morning, promoting the destabilization of the lower troposphere. This warming enabled moist air parcels to rise rapidly by early afternoon, eventually triggering convective development \citep{Banares2021}. By 05 UTC (13 LST), a discrete tornadic supercell became evident in HIMAWARI-9 AHI satellite imagery over Pampanga. This cell exhibited a pronounced overshooting top, spatially coincident with the tornado reports from  Barangay Candating in Arayat (Fig. \ref{fig4}a2 and \ref{fig4}b2). Further inspection using the Daytime Convection RGB product (Fig. \ref{fig4}b2) revealed signatures of a strong updraft, as indicated by the presence of small ice particles within convective cloud towers. Larger hydrometeors were observed within the thicker high-level cloud structures near the area of interest. The interaction of these ice-phase particles may have contributed to the observed lighting activity through ice-nucleation and other microphysical processes, occurring within the vigorous updraft of the tornadic supercell. These aspects will be discussed in further detail in subsequent sections.

To investigate the mechanism that facilitated convection initiation, we analyzed the synoptic environment using multiple constant-pressure levels, as illustrated in Figure \ref{fig5}. At 500-hPa (Fig. \ref{fig5}a-c), thorough inspection reveals that the proximity of TC Ewiniar induced subtle mid-level forcing for ascent especially as it moved away from the landmass, evident through lowering of the geopotential height contours. These height falls contribute to surface cyclogenesis and the generation of associated vorticity, which can promote atmospheric destabilization. Although the mid-level perturbation appeared relatively weak, it favored discrete cell development of discrete convective cells \citep{Schumann2010}. In contrast, stronger synoptic typically supports mixed or linear convective modes \citep{Bunkers2006,Dial2010}\footnote{In fact, the Oklahoma–Kansas tornado outbreak of 3 May 1999 was an example of weak-ill defined forcing \citep{Thompson2000}.}. A mid-level trough was also apparent which can demarcate the location of the advecting low-level moisture. However, the mid-tropospheric wind speeds remained modest, ranging from 10$-$15 knots (5$-$7 m s$^{-1}$).

At 700-hPa (Fig. \ref{fig5}d-f), slight decrease of geopotential heights was also observed in the vicinity of the event. These were accompanied by  precipitable water (PWAT) $>$ 55 kg m$^{-2}$ during the analysis period, likely due to the sustained influence and intensification of TC Ewiniar. Winds at this level were primarily westerly, as opposed to the northwesterly flow evident in the upper-levels, suggesting the presence of directional wind shear. These westerlies may have aided the enhancement of low-level southwesterlies across the area of interest. The 850-hPa level (Fig. \ref{fig5}g-i), representing the planetary boundary layer (PBL), showed pronounced warm and moist air advection associated with the SWM. The monsoon flow, characterized by southwesterlies drawn in by the circulation of TC Ewiniar, acted as a low-level jet (LLJ). Dewpoint temperatures $T_{\text{D}}$ $>$ 18 °C across much of Central Luzon, indicating a deep and high quality of low-level moisture (or boundary-layer moisture). This abundance of low-level moisture, combined with subtle mid-level forcing and strong surface heating during the morning hours, contributed to the thermodynamic instability. These ingredients are well-known pre-requisites for the initiation and maintenance of storm updrafts, including those found in supercells \citep{Pucik2015,Rasmussen2020}. Thus, the large-scale advective mechanism, such as the influence of a moving-away TC and its associated trough axis interacting with the SWM, can play a crucial role in establishing conditions favorable for tornadogenesis \citep{Doswell2001}.

\begin{figure*}[t]
\centering
\includegraphics[width=\textwidth]{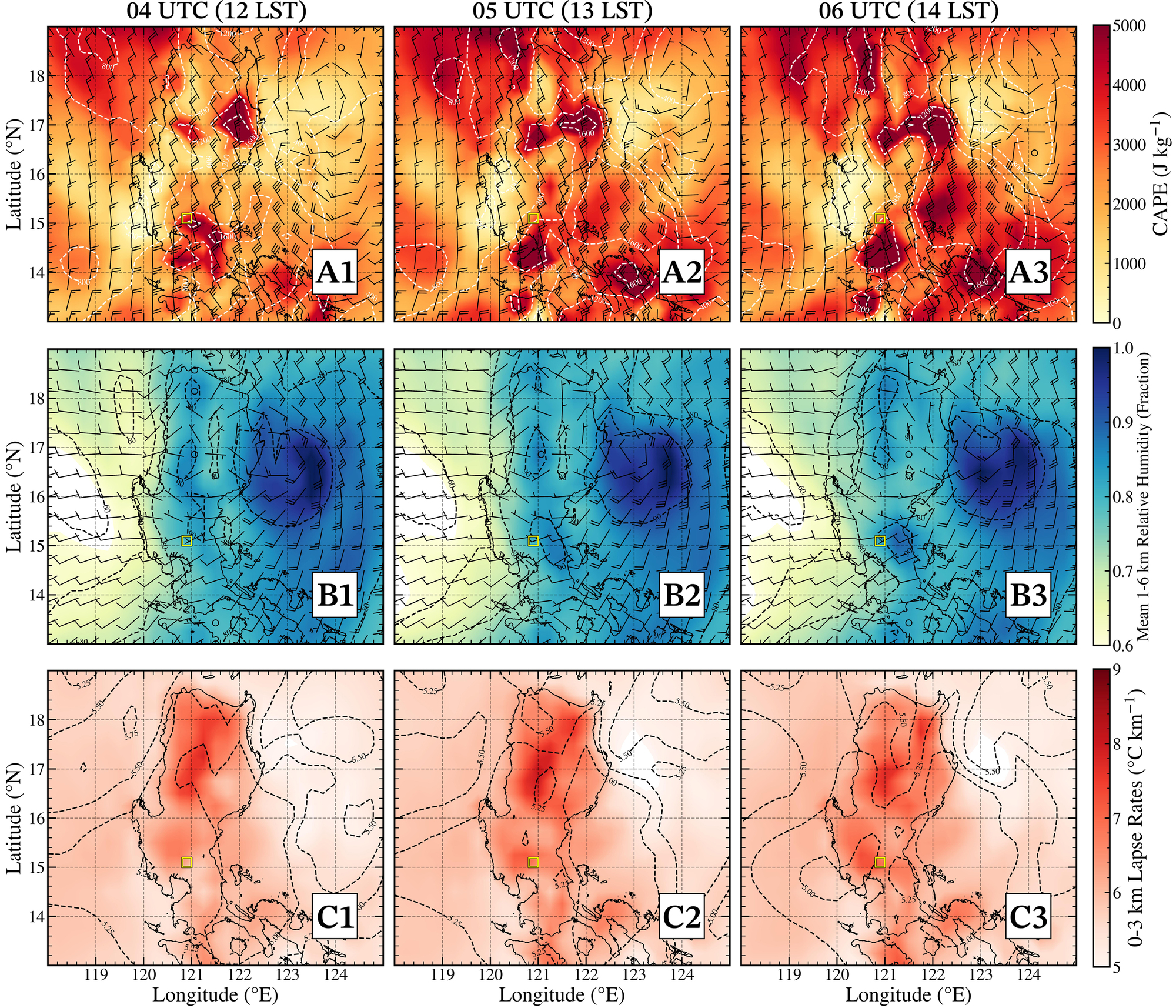}
\caption{Convective and Kinematic environment of the Luzon landmass before (1; 04 UTC / 12 LST), during (2; 05 UTC / 13 LST), and after (3; 06 UTC / 14 LST) the event. (a) CAPE (J kg$^{-1}$), 0-6 km Bulk Shear (kt), and WMAXSHEAR (white dashed; m$^{2}$ s$^{-2}$). (b) Average 1-6 km Relative Humidity (fraction), including Average 1-3 km Relative Humidity (black dashed; fraction), and 10 m Winds (kt). (c) Contours of 0-3 km Lapse Rates (°C km$^{-1}$) and 3-6 km Lapse Rates (black dashed; °C km$^{-1}$). The case area is demarcated in yellow box}
\label{fig8}
\end{figure*}

To support the findings from the HIMAWARI-9 AHI scans, several dynamic cross sections across the regions of interest are shown in Figure \ref{fig6}. In particular, negative values of the Brunt-Vaisala frequency squared ($N^2$) in Figure \ref{fig6}a are indicative of convective instability highlighting the presence of an unstable environment ($N^2$ $<$ 0). This instability was complemented by a convergence zone ($\nabla$F $<$ 0) along the storm’s initiation region (Fig. \ref{fig6}b). The convergence was likely enhanced by thermal winds i.e., slope and mountain-induced circulations, given the proximity of the event to Mt. Arayat. Such orographic influence may have caused the enhancement where northwesterly flow split around the mountain’s flanks and edges and may rejoin i.e., converge on the lee side, promoting lifting of warm, moist air parcels and initiating deep convection (Kalthoff et al. 2009, 2011; Kirshbaum et al. 2018). Further analysis shows negative vertical velocity anomalies reaching $\sim$$-$1 Pa s$^{-1}$ in the lower troposphere, particularly  between 1000 to 800 hPa pressure levels (Fig. \ref{fig6}c). These  upward vertical motions are crucial for sustaining deep convection and updrafts \citep{Wakimoto2004}. In addition, Figure \ref{fig6}d indicates the coexistence of positive values of both relative and absolute vorticity along the area of interest, suggesting enhanced rotational potential. These processes may have been further modulated by local topography, especially through thermally induced plain-to-mountain circulations, in conjunction with moist southwesterly low-level inflow and northwesterly upper-level winds, as discussed in the synoptic analysis.

\begin{figure*}[t]
    \centering
    \includegraphics[width=\textwidth]{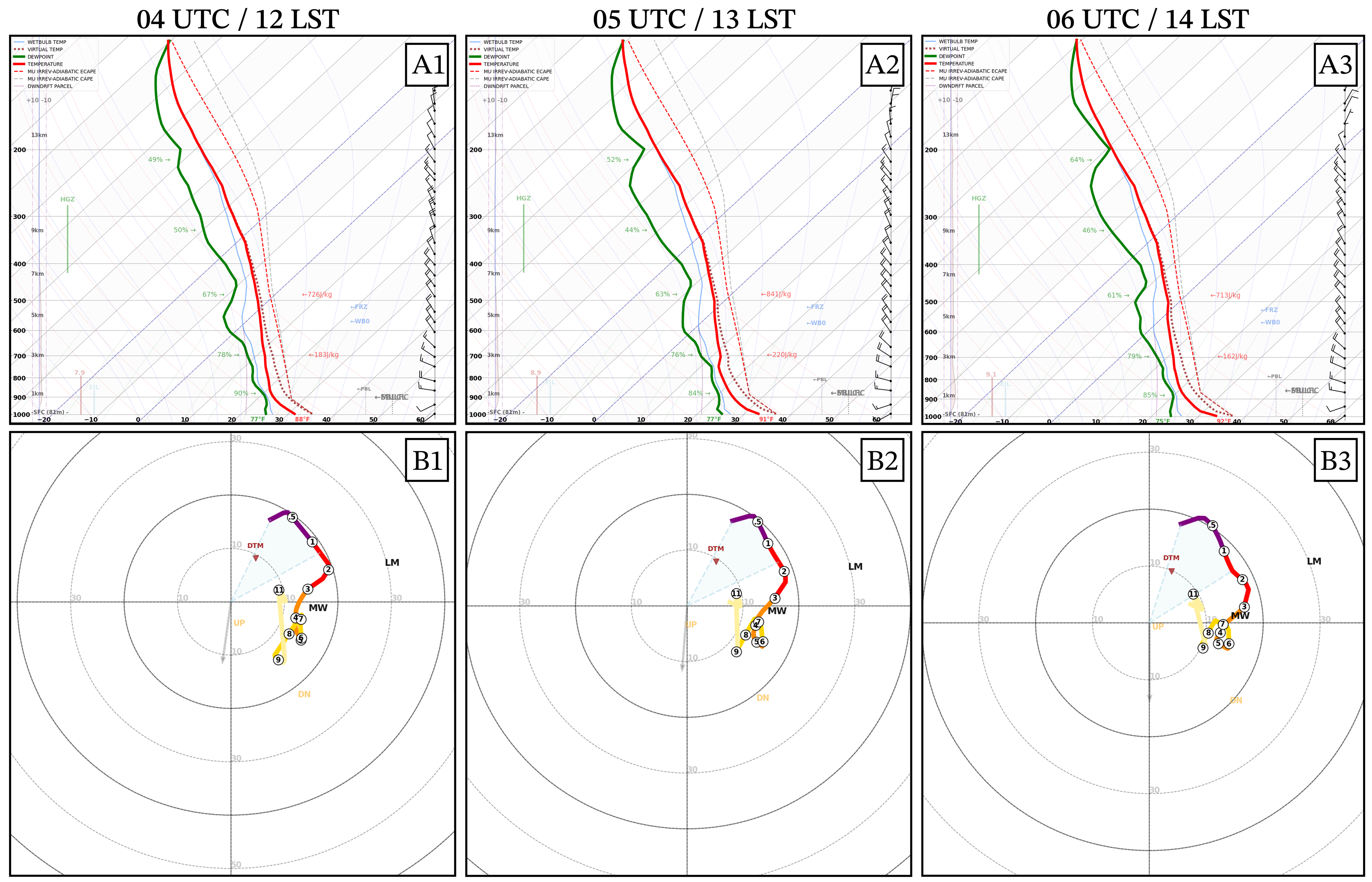}
    \caption{ERA5 Sounding Profile before (1; 04 UTC / 12 LST), during (2; 05 UTC / 13 LST), and after (3; 06 UTC / 14 LST) the event. (a) Thermodynamic profile along the area of interest. Annotations to the thermodynamic profiles include the SBLCL, PBL, MULFC, FRZ, and MUEL. (b) Storm-relative Hodographs associated with the thermodynamic profile above. Annotations to the wind profiles include the individual Bunkers’ Storm Motion, Deviant Tornado Motion, and Corfidi’s MCS components}
\label{fig9}
\end{figure*}

\subsection{Mesoscale Environment}

Analyzing the atmospheric conditions during the early morning hours is essential for understanding the background environment that influenced storm development later in the day. The pre-convective conditions are illustrated in Figures \ref{fig7}a and \ref{fig7}b using the 00 UTC (08 LST) sounding from DOST-PAGASA’s upper-air station in Tanay, Rizal, located approximately 87 km from the area of convective initiation. As a close-proximity profile, this sounding provides critical insight into the thermodynamic and kinematic state of the environment prior to storm development. 

As shown in Figure \ref{fig7}a, the initial thermodynamic condition was characterized by minimal temperature-dewpoint ($T$$-$$T_{\text{D}}$) spreads in the lowest 3 km AGL, with a deep moist layer extending up to 5 km. Relative humidity values exceeded 90\% in both the 1$-$3 km (RH$_{13}$) and 1$-$6 km (RH$_{16}$) layers. Two ‘warm nose’ inversions were observed above the 700-hPa level, which likely inhibited the lifting of surface-based parcels until sufficient moisture advection and diurnal heating cool these features. While overall lapse rates were also muted to start the day, the mid-level lapse rates reached $-$5.7 °C km$^{-1}$ due to the inversions between 3$-$6 km, resulting in a modest steepening of temperature gradients aloft. This results in an ample undiluted CAPE of 1484 J kg$^{-1}$, along with a low-level CAPE of 196 J kg$^{-1}$. 

\begin{table*}[h!t!]
\caption{Sounding-derived measurements associated with the tornadic supercell event.}
    \centering
    \begin{tabular}{lllll}
    \hline\hline
    Parameter & 00 UTC Obs & 04 UTC ERA5 & 05 UTC ERA5 & 06 UTC ERA5\\
    \hline
    CAPE & 1484 J kg$^{-1}$ & 3255 J kg$^{-1}$ & 3751 J kg$^{-1}$ & 2925 J kg$^{-1}$ \\
    CAPE$_{03}$ & 196 J kg$^{-1}$ & 183 J kg$^{-1}$ & 220 J kg$^{-1}$ & 162 J kg$^{-1}$ \\
    CAPE$_{06}$ & 547 J kg$^{-1}$ & 726 J kg$^{-1}$ & 841 J kg$^{-1}$ & 713 J kg$^{-1}$ \\
    ECAPE & 864 J kg$^{-1}$ & 2084 J kg$^{-1}$ & 2500 J kg$^{-1}$ & 1820 J kg$^{-1}$ \\
    CIN & 0 J kg$^{-1}$ & 0 J kg$^{-1}$ & 0 J kg$^{-1}$ & 0 J kg$^{-1}$ \\
    LCL & 40 m & 782 m & 997 m & 1228 m \\
    LFC & 40 m & 782 m & 997 m & 1228 m \\
    LR$_{03}$ & 5.11 $^{\circ}$C km$^{-1}$ & 7.09 $^{\circ}$C km$^{-1}$ & 7.71 $^{\circ}$C km$^{-1}$ & 7.96 $^{\circ}$C km$^{-1}$ \\
    LR$_{36}$ & 5.71 $^{\circ}$C km$^{-1}$ & 5.50 $^{\circ}$C km$^{-1}$ & 5.43 $^{\circ}$C km$^{-1}$ & 5.40 $^{\circ}$C km$^{-1}$ \\
    RH$_{13}$ & 0.90 / 90$\%$ & 0.82 / 82$\%$ & 0.74 / 74$\%$ & 0.84 / 84$\%$ \\
    RH$_{16}$ & 0.98 / 98$\%$ & 0.73 / 73$\%$ & 0.73 / 73$\%$ & 0.67 / 67$\%$ \\
    BWD$_{01}$ & 13 kt / 6.7 m s$^{-1}$ & 8 kt / 4.1 m s$^{-1}$ & 7 kt / 3.6 m s$^{-1}$ & 8 kt / 4.1 m s$^{-1}$ \\
    BWD$_{03}$ & 15 kt / 7.7 m s$^{-1}$ & 14 kt / 7.2 m s$^{-1}$ & 15 kt / 7.7 m s$^{-1}$ & 18 kt / 9.3 m s$^{-1}$ \\
    BWD$_{06}$ & 23 kt / 11.8 m s$^{-1}$ & 22 kt / 11.3 m s$^{-1}$ & 22 kt / 11.3 m s$^{-1}$ & 22 kt / 11.3 m s$^{-1}$ \\
    BWD$_{13}$ & 12 kt / 6.2 m s$^{-1}$ & 8 kt / 4.1 m s$^{-1}$ & 9 kt / 4.6 m s$^{-1}$ & 10 kt / 5.1 m s$^{-1}$ \\
    BWD$_{16}$ & 18 kt / 9.2 m s$^{-1}$ & 18 kt / 9.2 m s$^{-1}$ & 17 kt / 8.7 m s$^{-1}$ & 16 kt / 8.2 m s$^{-1}$ \\
    B2K$_{\text{RM}}$ & 5 kt / 2.5 m s$^{-1}$ & 10 kt / 5.1 m s$^{-1}$ & 11 kt / 5.6 m s$^{-1}$ & 13 kt / 6.7 m s$^{-1}$ \\
    V$_{SR}$ & 13 kt / 6.6 m s$^{-1}$ & 17 kt / 8.7 m s$^{-1}$ & 17 kt / 8.7 m s$^{-1}$ & 19 kt / 9.7 m s$^{-1}$ \\
    SRH$_{500}$ & 11 m$^{2}$ s$^{-2}$ & 16 m$^{2}$ s$^{-2}$ & 19 m$^{2}$ s$^{-2}$ & 27 m$^{2}$ s$^{-2}$ \\
    SRH$_{01}$ & 23 m$^{2}$ s$^{-2}$ & 46 m$^{2}$ s$^{-2}$ & 40 m$^{2}$ s$^{-2}$ & 50 m$^{2}$ s$^{-2}$ \\
    SRH$_{03}$ & 73 m$^{2}$ s$^{-2}$ & 89 m$^{2}$ s$^{-2}$ & 88 m$^{2}$ s$^{-2}$ & 100 m$^{2}$ s$^{-2}$ \\
    SRH$_{13}$ & 50 m$^{2}$ s$^{-2}$ & 42 m$^{2}$ s$^{-2}$ & 48 m$^{2}$ s$^{-2}$ & 49 m$^{2}$ s$^{-2}$ \\
    $\omega_{s500}$ & 0.004 s$^{-1}$ & 0.004 s$^{-1}$ & 0.004 s$^{-1}$ & 0.006 s$^{-1}$ \\
    $\widetilde{\omega_s}$$_{500}$ & 0.56 / 56\% & 0.74 / 74\% & 0.80 / 80\% & 0.86 / 86\% \\
    $\omega_{s01}$ & 0.004 s$^{-1}$ & 0.005 s$^{-1}$ & 0.004 s$^{-1}$ & 0.005 s$^{-1}$ \\
    $\widetilde{\omega_s}$$_{01}$ & 0.52 / 52\% & 0.86 / 86\% & 0.87 / 87\% & 0.88 / 88\% \\
    $\omega_{\text{max}}$ & 0.006 s$^{-1}$ & 0.007 s$^{-1}$ & 0.006 s$^{-1}$ & 0.007 s$^{-1}$ \\
    $\widetilde{\omega}$$_{\text{max}}$ & 0.68 / 68\% & 0.99 / 99\% & 0.97 / 97\% & 0.99 / 99\% \\
    CA & 134$^{\circ}$ & 122$^{\circ}$ & 116$^{\circ}$ & 105$^{\circ}$ \\
    DTM & 8 kt / 4.1 m s$^{-1}$ & 4 kt / 2.0 m s$^{-1}$ & 5 kt / 2.5  m s$^{-1}$ & 5 kt / 2.5  m s$^{-1}$ \\
    $\zeta_{\text{LLM}}$ (max) & 0.012 s$^{-1}$ (0.017 s$^{-1}$) & 0.011 s$^{-1}$ (0.015 s$^{-1}$) & 0.010 s$^{-1}$ (0.014 s$^{-1}$) & 0.010 s$^{-1}$ (0.013 s$^{-1}$)\\
    WMAXSHEAR & 616.58 m$^{2}$ s$^{-2}$ & 913.17 m$^{2}$ s$^{-2}$ & 980.28 m$^{2}$ s$^{-2}$ & 864.28 m$^{2}$ s$^{-2}$ \\
    \hline
    \end{tabular}
    \label{table:3}
\end{table*}

The associated hodograph in Figure \ref{fig7}b illustrates the veering winds with height, though upper-level flow was predominantly westerly to northwesterly. This wind profile induced only minimal clockwise curvature in the hodograph shape, as compared in tornadic storms with large, looping shapes commonly associated with tornadic environments featuring strong southerly i.e., backed flow or southwesterly low-level winds \citep{Nixon2022}. The mean V$_{SR}$ in the lowest 1 km was $<$ 10 m s$^{-1}$, suggesting that initial convective updrafts may have been susceptible to entrainment dilution under a moist environment as convective clouds may only realize around 58\% of the available instability under such conditions \citep{Peters2019a,Peters2019b}\footnote{Known as ECAPE-to-CAPE ratio, $\Tilde{E}$.}. In addition, the background vorticity was not fully streamwise, with streamwise fractions of only 52–56\% in the lowest 500 m and 1 km layers suggesting that the earliest convective attempts may have been inefficient, with updrafts not only ingesting crosswise vorticity but also struggling to maintain buoyancy during their formative stages. Despite these limitations, the broader environment still possessed ingredients supportive of more organized convection later in the day. DLS of approximately 23 kts ($\sim$12 m s$^{-1}$) and SRH$_{03}$ of 73 m$^{2}$ s$^{-2}$ fall within the parameter space associated with supercell-supporting environments \citep{RasmussenBlanchard1998}, although storms may take time to congeal and mature as they experience more dilution, aside on the lack of V$_{SR}$, due to the DLS $>$ 20 kt in place tilting the initial updrafts \citep{Lebel2023}. In addition, the enhancement of low-level kinematics may have to rely on other mesoscale features, such as the influence of nearby terrain (e.g., Mt. Arayat) or interactions between developing storm cells.

Furthermore, the mesoscale environment associated with the tornadic event was assessed using ERA5 reanalysis data, as illustrated in Figure \ref{fig8}. As depicted in Figures \ref{fig8}a1-a3, CAPE measurements $>$ 3000 J kg$^{-1}$ were depicted suggesting ample instability in the atmosphere. Modest DLS, represented by the 0$-$6 km bulk shear, ranged from  20$-$25 kts recorded across Central Luzon, sufficient to support organized convective storms. These conditions also produced high values of WMAXSHEAR parameter $>$ 900 m$^{2}$ s$^{-2}$ during the analysis period, representative of the storm’s severity. The combination of high CAPE and DLS is commonly observed in tornadic environments during spring-season severe weather outbreaks. In such regimes, the shear vector often exhibits pronounced clockwise curvature with height, especially within the lower levels and forming a ‘sickle shaped’ profile that is favorable for the development of tornadic supercells \citep{Weisman1986,Thompson2000,Coffer2020,Nixon2022}. The DLS vectors retrieved from ERA5 also aligned well with the observed wind profiles from proximity sounding discussed earlier. Notably, the shear axis appeared oriented perpendicular to the initiating boundary, which may have been defined by localized terrain gradients or differential heating. This configuration likely contributed to the storm’s semi-discrete or even discrete convective mode, a structural feature often associated with tornadic development.

\begin{figure*}[t]
\centering
\includegraphics[width=0.8\textwidth]{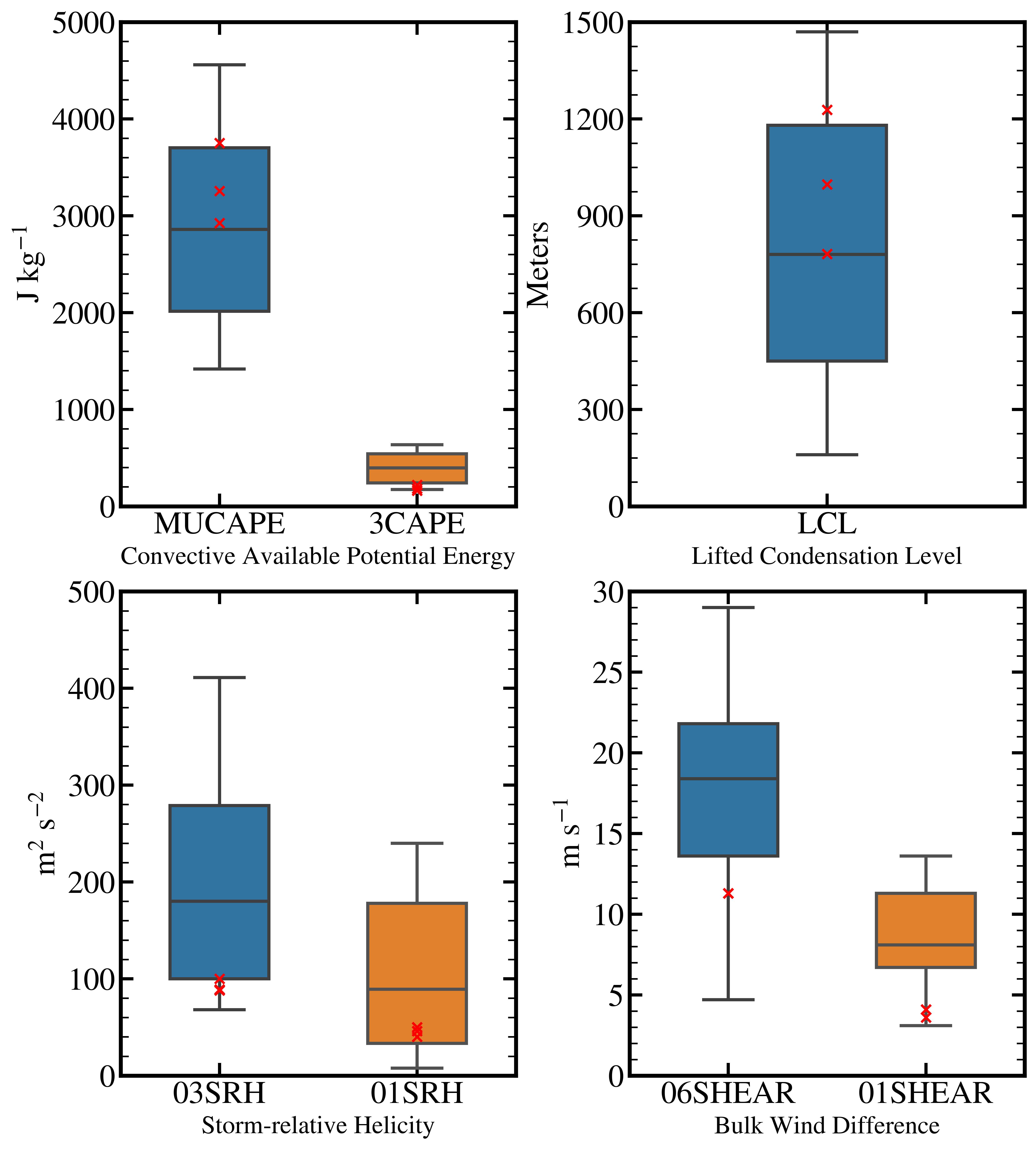}
\caption{Comparison of the parameters of the Candating, Arayat tornadic environment (red X) calculated with the ERA5 model soundings in Fig. 8 with the U. S. climatology for a supercellular tornado environment given by \citet{RasmussenBlanchard1998} and \citet{Thompson2003}. Parameters include MUCAPE (J kg$^{-1}$), 0-3 km CAPE (J kg$^{-1}$), LCL (m), both 0-3 km and 0-1 km SRH (m$^{2}$ s$^{-2}$), and both 0-6 km and 0-1 km Bulk Shear (m s$^{-1}$)}
\label{fig10}
\end{figure*}

In Figures \ref{fig8}b1-b3, the low-level and mid-level RH fields show a moistening of the environmental profile, with RH $>$ 70\% across the Candating, Arayat region (highlighted in the yellow box). To the east of the storm initiation zone, layers above cloud base exhibited RH $>$ 90\%, which may have contributed to the maintenance of additional convective development. This moist environment was superimposed on steep low-level temperature lapse rates exceeding $>$ $-$6 °C km$^{-1}$ within the lowest 3 km, as illustrated in Figures \ref{fig8}c1-c3, from 04$-$06 UTC (12$-$14 LST). The pronounced near-surface temperature gradient enabled buoyant air parcels to precipitously rise, enhancing the overall instability within the area of interest. In the mid-troposphere, lapse rates approached $-$5.5 °C km$^{-1}$, further supporting high CAPE values discussed earlier. These conditions indicate that thermodynamic instability was strongly enhanced by the combination of surface heating and the advection of moist air via the prevailing southwesterly flow. This alignment of steep lapse rates and elevated RH at both low- and mid-level facilitated increased CAPE and provided sufficient buoyancy to sustain robust convective updraft \citep{Pucik2015}. 

\begin{figure*}[t]
\centering
\includegraphics[width=\textwidth]{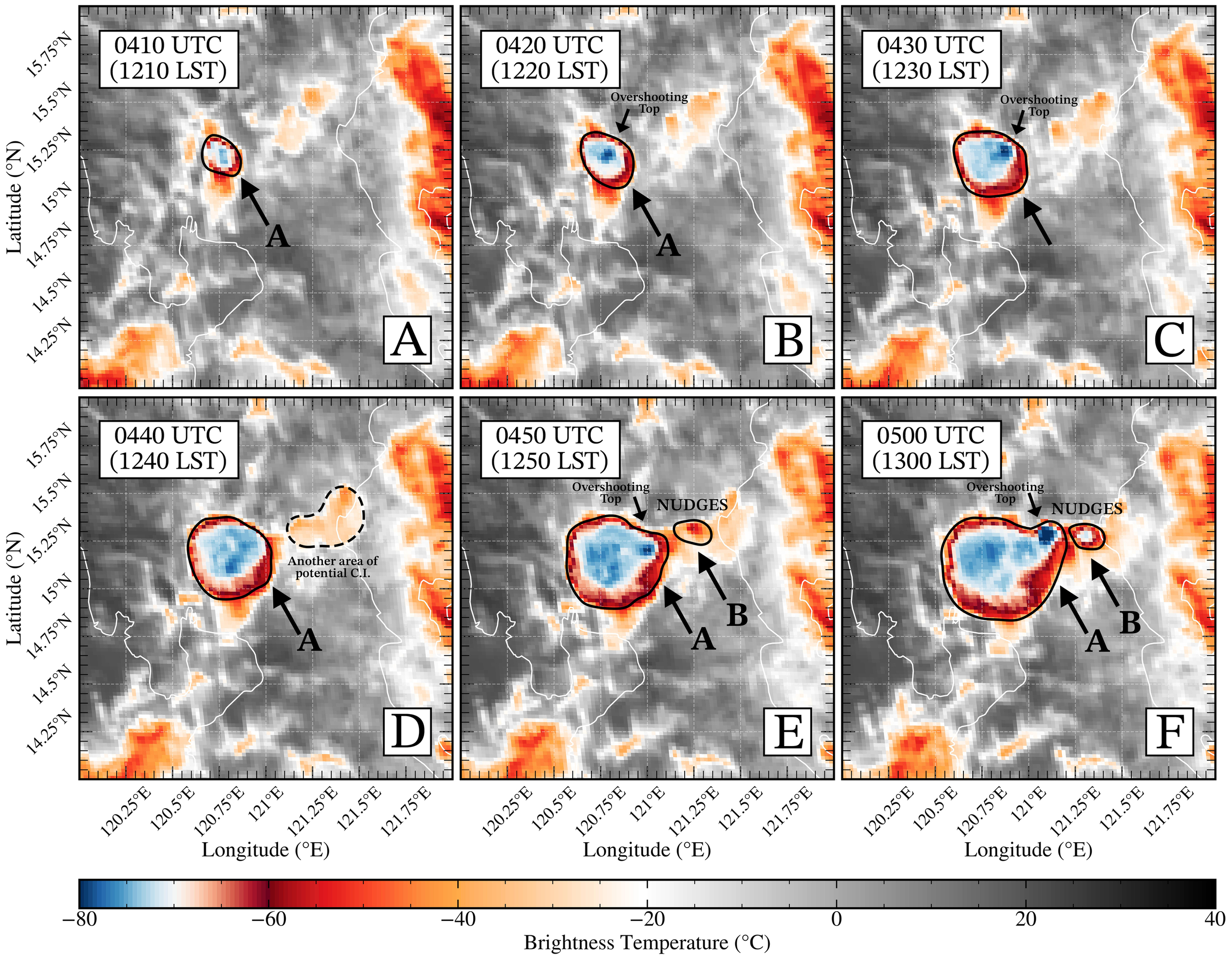}
\caption{HIMAWARI-9 AHI 10.4 μm BT depicting storm evolution of the 27 May 2024 tornadic supercell from initial to mature phase. Each scans are every 10 minutes, starting at 0410 to 0500 UTC (1210 to 1300 LST; a-f)}
\label{fig11}
\end{figure*}

To further examine the evolving thermodynamic and kinematic environment, Figure \ref{fig9} presents the derived model sounding and hodographs at 04, 05, and 06 UTC (12, 13, and 14 LST), near the tornado but outside the convectively contaminated regions. Compared to the 00 UTC (08 LST) observed pre-convective sounding, the model sounding (Fig. \ref{fig9}a1-a3) exhibit a subtle $inverted-V$ profile characterized by a well-mixed near-surface layer, the absence of a significant inhibition layer, and a relatively low lifting condensation level (LCL) ranging from 700$-$1000 m AGL. Additionally, the profiles indicate a moist PBL with RH$_{13}$ $>$ 80\% and steep low-level lapse rates throughout the entire event duration, approaching $-$7 °C km$^{-1}$. Although the moisture quality between the lower and mid-tropospheric levels (represented by RH$_{16}$) decreased by $\sim$20\% due to the eventual surface heating, the thermodynamic environment remained highly favorable for convective development. The well-mixed and buoyant conditions allowed warm air parcels to rise efficiently, resulting in high undiluted MUCAPE values exceeding 3000 J kg$^{-1}$. Notably, at 05 UTC (13 LST), coinciding with the tornado occurrence in Candating, Arayat, MUCAPE peaked at 3751 J kg$^{-1}$, representing a threefold increase in instability compared to earlier soundings. This substantial increase in instability highlights a thermodynamic environment that strongly favored surface-based deep convection \citep{Matsui2016}.

\begin{figure*}[t]
\centering
\includegraphics[width=0.934\textwidth]{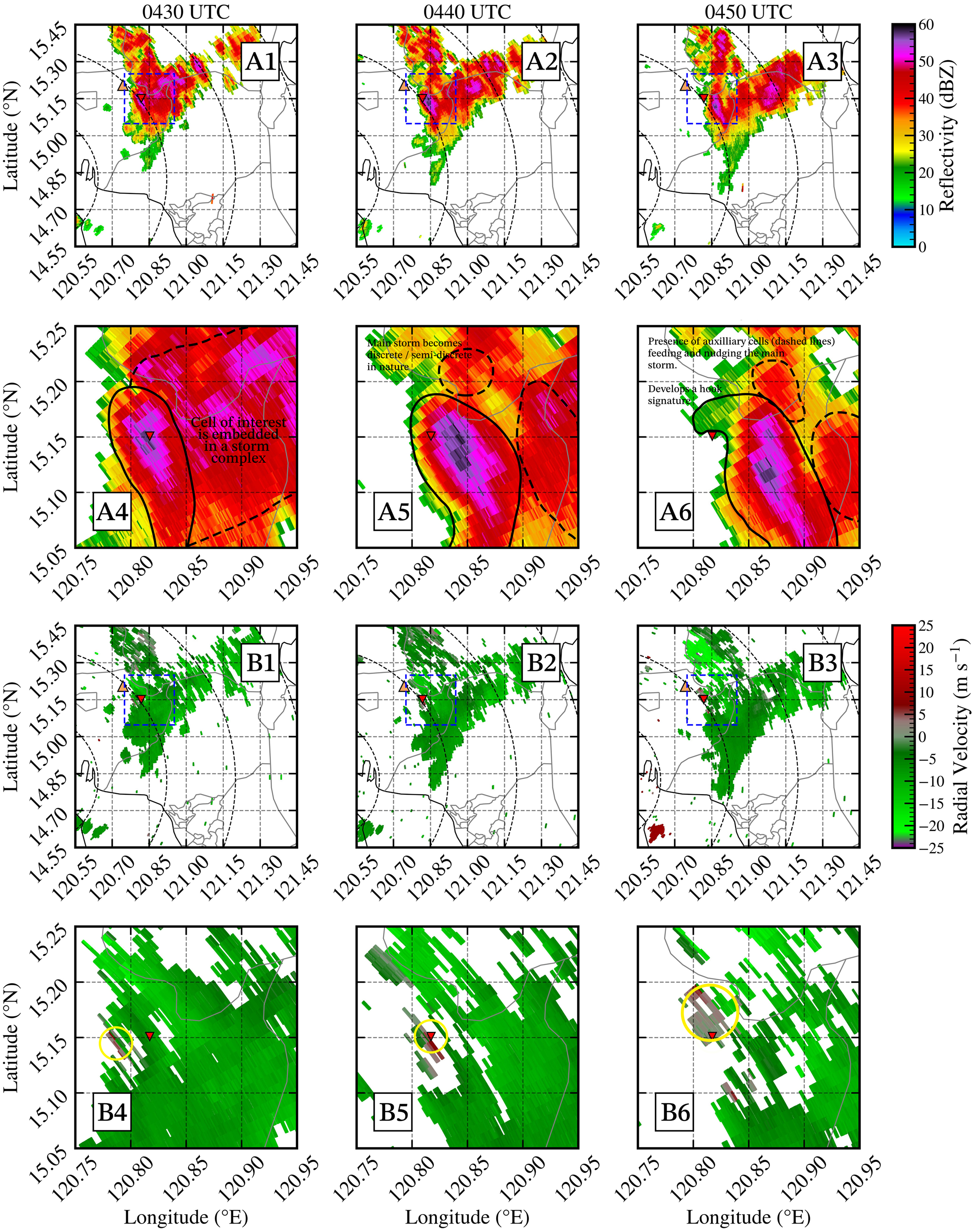}
\caption{S-SUB lowest elevation scans of the tornadic supercell for every 10 minutes. Radar variables include (a) Reflectivity (dBZ) and (b) Radial Velocity (m s$^{-1}$). Time stamps start from 0430$-$0450 UTC (1230$-$1250 LST). Both a4$-$a6 and b4$-$b6 sub-figures are ‘zoomed in’ radar scans to their respective time and radar products. Inverted red triangle is the location of the Candating, Arayat, while the orange triangle is the location of Mt. Arayat. Yellow circle is the inherent velocity couplet associated with the tornado.}
\label{fig12}
\end{figure*}

\begin{figure*}[t]
\centering
\includegraphics[width=\textwidth]{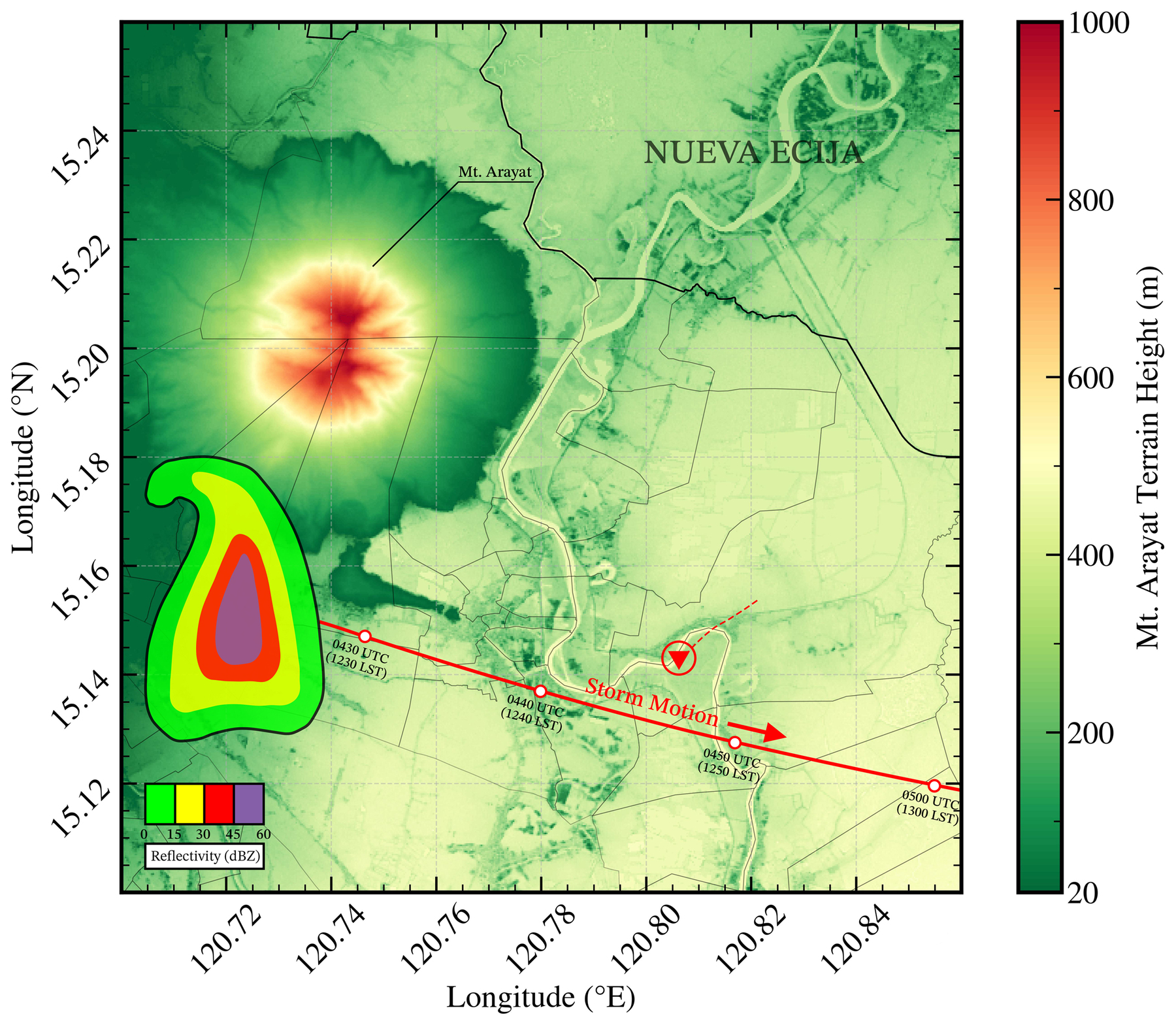}
\caption{Elevation map of Arayat, Pampanga and nearby municipalities. Hypothetical low-level structure of the Candating, Arayat supercell on radar is included. Radar reflectivity is color shaded in dBZ. The solid red line is the supercell’s track accompanied by the respective radar scan time stamps, while the broken red line is the $\sim$2 km tornado path with an enclosed red triangle.}
\label{fig13}
\end{figure*}

One defining feature of the observed and modeled hodographs associated with this event was their unusual orientation (Fig. \ref{fig7}b and Fig. \ref{fig9}b1-b3). The hodograph exhibited a clockwise rotation exceeding three octants from the classical configuration, indicating a unique shear environment. This wind profile produced DLS vectors that supported convective storms with southeastward motion, with right-moving (RM) supercells veering further eastward. Such hodograph shape is characteristic of the warm-season North American Monsoon (NAM) regime \citep{Adams1997}. \citet{Blanchard2011} noted that these kinds of hodographs can favor hook echo signatures in the northwestern quadrant of supercells. In this case, RM storm motions ranged from 005°$-$000° at speeds of 10–15 kt with a vector-averaged mean motion of $\sim$320° at 17 kt (8.7 m s$^{-1}$). Variability in the motion of the tornadic supercell was likely influenced by storm-storm and storm-outflow interactions \citep{Zeitler2005}, and potentially by local terrain effects. Notably, V$_{SR}$ increased to $\sim$17 kt compared to the earlier soundings, which helped mitigate the negative impacts of entrainment dilution on supercell’s updraft. A fractional entrainment of $\sim$64\% was computed, indicating that the storm’s updraft retained a substantial portion of the undiluted CAPE. This suggests that more than half of the available convective energy was effectively converted into  updraft kinetic energy \citep{Romps2010}. 

\begin{figure*}[t]
\centering
\includegraphics[width=\textwidth]{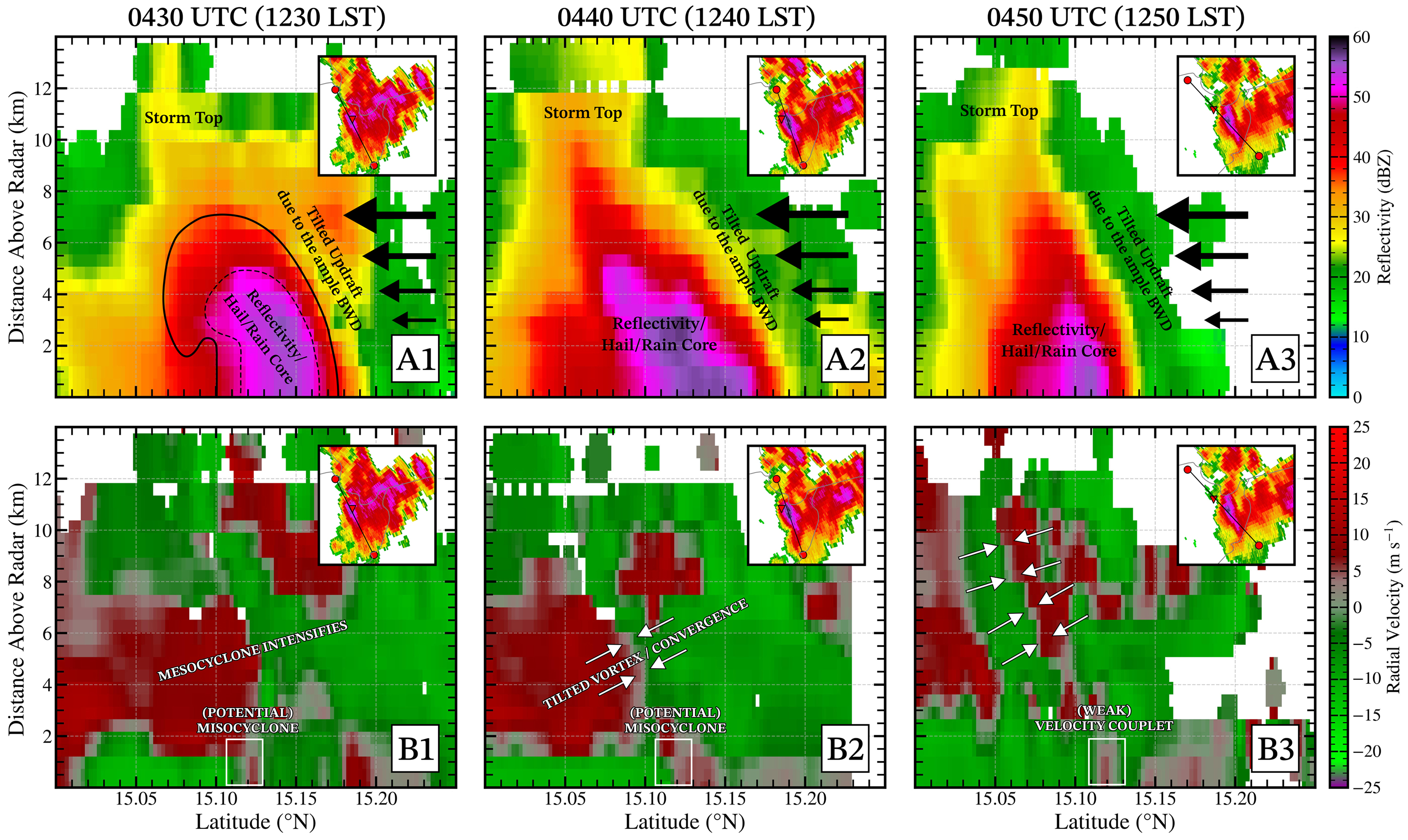}
\caption{Vertical Cross Section of S-SUB (a) Reflectivity (dBZ) and (b) Radial Velocity (m s$^{-1}$) along different points in the inset axes at 0430$-$0450 UTC (1230$-$1250 LST). The cross section is oriented parallel to the storm’s motion to highlight its internal structure.}
\label{fig14}
\end{figure*}

Additional kinematic parameters derived from the model storm-relative hodographs further support a favorable tornadic environment. Although the ambient near-surface vorticity remained relatively weak ($\sim$0.004 s$^{-1}$), it became increasingly aligned with the storm-relative flow near the time of tornadogenesis at 05 UTC (13 LST), with \textgreek{ῶ}$_{\text{s500}}$ $>$ 80\% at critical angles (CA) $<$ 120°. In the lowest 1 km layer, horizontal vorticity also remained modest in magnitude but was highly streamwise throughout the analysis period, with \textgreek{ῶ}$_{\text{s01}}$ $>$ 85\% with maximum 0$-$1 km streamwiseness $>$ 95\%. High streamwiseness, particularly from the surface to the cloud base, are crucial for tornadogenesis as they allow environmental vorticity to be more readily ingested, tilted, and stretched into the vertical by the storm’s updraft. Although the model soundings did not resolve enhanced low-level shear (LLS $<$ 10 kt), topographically induced modifications were likely present. Specifically, terrain-driven processes such as plain-to-mountain flow and lee-side convergence may have increased ambient vorticity and enhanced LLS which unfortunately, was not captured by the model soundings as it remained weak. These processes were associated with the supercell’s motion toward terrain-induced cyclonic vorticity anomalies on the western slope of Mt. Arayat. Southwesterly flow on the windward side and channeling of northwesterlies toward the lee likely contributed to low-level convergence \citep{Kirshbaum2018}. Such terrain influences are consistent with numerical simulation by \citet{Markowski2010}, which demonstrated that lee-side convergence enhances both of CAPE and SRH near the lee slope, thereby facilitating the rapid spin-up of low-level rotation and intensification of the supercell. Despite the weak streamwise vorticity in the lowest kilometer, the CAPE confined to the 0$-$3 km layer (CAPE$_{03}$) compensated for this deficiency, with vertical vorticity in the low-level mesocyclone exceeding 0.01 s$^{-1}$ throughout the analysis period adequate for supercells to consummate a tornado \citep{Coffer2023}.

The composite parameter WMAXSHEAR, which combines instability and DLS, exceeded m$^2$ s$^{-2}$ near the time of the tornadic event, representing a significant increase from the morning value of 616 m$^2$ s$^{-2}$. This magnitude falls well within the climatological range associated with environments favorable for severe convection and tornadogenesis (Taszarek et al. 2020). As illustrated in Figure \ref{fig10}, model-derived sounding parameters were compared to established climatological baseline outlined in \citet{RasmussenBlanchard1998} and \citet{Thompson2003}. Notably, the kinematic parameters such as DLS and SRH across various layers, fell between the lower bound and first quartile (Q1) of typical tornadic supercell environments. Meanwhile, thermodynamic variables such as MUCAPE and LCL were positioned near the third quartile (Q3), suggesting ample instability and a low storm base conducive to tornadogenesis. These findings indicate that the background environment was conditionally supportive of discrete tornadic supercells. Moreover, additional mesoscale influences, particularly terrain-induced modifications, likely enhanced the ambient kinematics and buoyant instability available for storm development. However, the persistence, spatial extent, and tornadic nature of the supercell may also be attributed to internal storm-scale dynamics, which will be addressed in the following subsection. A summary of key thermodynamic and kinematic indices derived from both the observed and ERA5 model sounding is provided in Table \ref{table:3} for comparison and reference.

\subsection{Storm Development and Life Cycle}

\subsubsection{Supercell Dynamics}

The temporal evolution of the tornadic storm is illustrated in Figure \ref{fig11} using 10-minute interval scans from HIMAWARI-9 between 0410$-$0500 UTC (1210$-$1300 LST). While satellite data prior to 0410 UTC and beyond 0500 UTC are not visualized, the relevant information outside this period is discussed to provide a descriptive context of the storm’s life cycle.

\begin{figure*}[t]
\centering
\includegraphics[width=\textwidth]{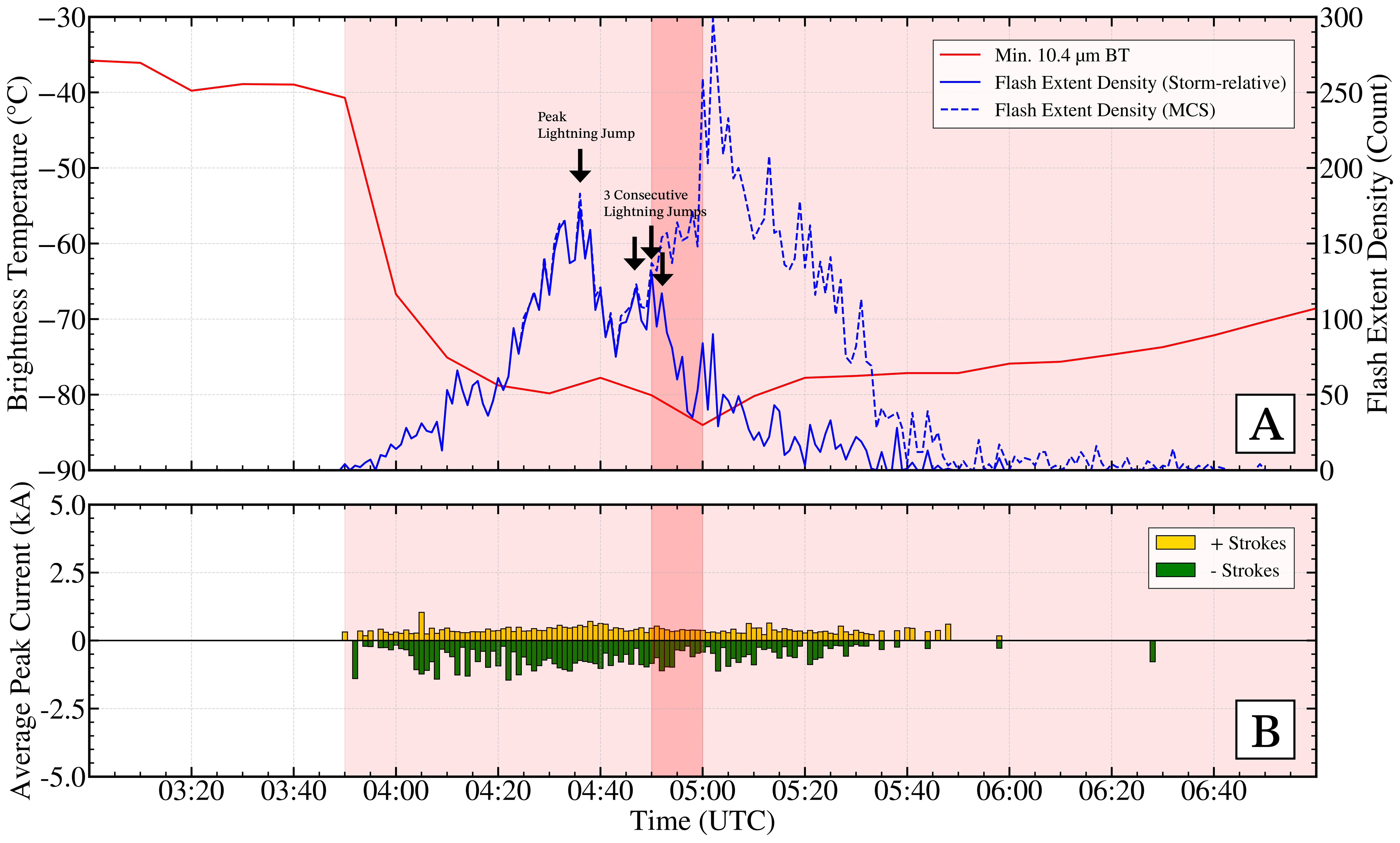}
\caption{Time series of (a) Minimum 10.3 μm brightness temperature in the storm (red; °C), maximum flash extent density per minute (blue; counts or flashes min$^{-1}$) for: storm-relative (solid line) and over the entire MCS (dashed lines). (b) Average Peak Current (kA) per minute of recorded Positive (yellow) and Negative Strokes (green) within the supercell. Highlighted in the time series are the estimated severe thunderstorm occurrence (light red) and tornadic event (dark red).}
\label{fig15}
\end{figure*}

At 03 UTC (11 LST), initial signs of convection were evident in the form of agitated cumulus clouds developing across Central Luzon. This development was driven by sustained surface heating within a moist environment, as previously described in the thermodynamic discussion. The rising parcels of warm, moist air were further influenced by southwesterly flow at the surface and northwesterlies aloft, which intersected the topography of Mt. Arayat. This orographic interaction generated a zone of lee-side convergence; an important mesoscale trigger for deep convection and supercell initiation. This development coincides with the thunderstorm advisory issued by DOST-PAGASA’s Regional Division located in the National Capital Region (PAGASA-NCR PRSD) at 0307 UTC (1107 LST) citing thunderstorms with strong winds and lightning across much of the GMMR is possible over the next few hours. The advisory covered several key provinces, including Bataan, Pampanga, Bulacan, Cavite, and Batangas.  

Between 0400 and 0410 UTC (1200$-$1210 LST), a discrete thunderstorm (hereafter Storm “A”) initiated in the vicinity of the Arayat area. As shown in Figure \ref{fig11}a-b, the storm rapidly intensified and expanded in spatial extent, developing a distinct overshooting top (OT) with a brightness temperature (BT) $\sim$$-$78 °C. By 0430 UTC (1230 LST; Fig. \ref{fig11}c), Storm “A” had entered its mature phase, maintaining a persistent OT feature leading up to the time of tornado initiation at 0450 UTC (1250 LST; Fig. \ref{fig11}d). In response to the rapidly developing convection, DOST-PAGASA’s NCR PRSD issued a thunderstorm advisory at 0445 UTC (1245 LST), reporting moderate to heavy rainfall, frequent lightning, and outflow winds affecting areas of Pampanga, particularly Candaba, Santa Ana, and Arayat. The advisory further indicated the likelihood of continued convective activity over the next two hours. Storm “A” exhibited a gradual eastward propagation, consistent with the orientation of the DLS vector described earlier. Between 0440$-$0500 UTC (1240$-$1300 LST; Fig. \ref{fig11}d-f), the parent storm exhibited extremely cold cloud tops indicative of strong vertical development and OT, with BT between $-$80 and $-$84 °C. 

During this period, an auxiliary convective cell (hereafter referred to as cell “B”) quickly developed and intensified to the east of the main storm cell between 0440 and 0450 UTC (1240$-$1250 LST; Fig. \ref{fig11}d-e). This secondary cell remained intact as Storm “A” reached the peak of its life cycle. This persistence may be attributed to an unimpeded moisture supply in the region east of the parent cell. Given their spatial proximity, it is plausible that cell ”B” contributed to storm maintenance by mitigating entrainment effects (due to lack of storm-relative inflow) by achieving inflow-outflow balance and providing ample storm ventilation. These observed storm interactions are consistent with recent findings in the literature. \citet{Fischer2023} and \citet{Nixon2024} documented that tornado- and hail-producing supercells are often accompanied by auxiliary convective cells that eventually nudges, feeds (i.e., partially merging), and merges, constructively (i.e., augmentation).

To complement the satellite-based analysis and evaluate the influence of auxiliary cell interactions, radar scans from the S-SUB were analyzed. Figure \ref{fig12} presents the evolution of the tornadic supercell from 0430$-$0450 UTC (1230$-$1250 LST) at 10-minute intervals, with both reflectivity and radial velocity scans. At the location of the observed damage, the lowest radar elevation angle (0.5°) corresponded to an approximated beam height of 550 m AGL. Reflectivity scans (Fig. \ref{fig12}a1-a3; zoomed in version is Fig. \ref{fig12}a4-a6) reveal a complex of convective storms. The  main storm cell, outlined in black, exhibited maximum reflectivities from 55$-$59 dBZ, coinciding with the satellite-identified OTs (BT $<$ $-$70 °C), indicative of persistent and vigorous updraft activity. Additionally, several smaller semi-discrete convective elements (dashed outline) were located predominantly along the left and rear flanks of the parent storm. These auxiliary cells are consistent with satellite-observed cell ‘B’ and align with the storm-relative configuration described in \citet{Nixon2024}, which suggests that such cells, forming 20$-$30 minutes prior tornadogenesis, can enhance inflow via generated outflow boundaries. This process may contribute to achieving inflow-outflow balance by decelerating pre-tornadic vertical advection; a known precursor to tornadogenesis under weak storm relative inflow \citep{Dowell2002}. 

\begin{table}[t!]
\caption{Overall lightning characteristics of the storm cell.}
\resizebox{\columnwidth}{!}{%
    \centering
    \begin{tabular}{ll}
    \hline\hline
    Storm cell duration & $\sim$2:10 hr \\
    \hline\hline
    $-$CG Flashes & 1604 flashes \\
    $-$CG Flashes with peak currents $<$ $-$10 kA & 413 flashes \\
    $+$CG Flashes & 4328 flashes \\
    $+$CG Flashes with peak currents $>$ 10 kA & 271 flashes \\
    Average peak current for $-$CG (kA) (median) & $-$7910.23 kA ($-$3841.51 kA) \\
    Average peak current for $-$CG ($<$ $-$10 kA) (median) & $-$20319.87 kA ($-$16927.05 kA) \\
    Average peak current for $+$CG (kA) (median) & 4290.61 kA (2787.04 kA) \\
    Average peak current for $+$CG ($>$ 10 kA) (median) & 19128.54 kA (14185.73 kA) \\
    Maximum Total Flash Rate (min$^{-1}$) & 173 flashes min$^{-1}$ \\
    Maximum $-$CG Rate (min$^{-1}$) & 133 flashes min$^{-1}$ \\
    Maximum $+$CG Rate (min$^{-1}$) & 59 flashes min$^{-1}$ \\
    \hline
    \end{tabular}
}
\label{table:4}
\end{table}

A key feature observed in the radial velocity scans was the emergence of a velocity couplet between 0440 and 0450 UTC (1240$-$1250 LST), marked in Figure \ref{fig12}c2$-$c3 and Figure \ref{fig12}d2$-$d3 (marked as an inverted red triangle and yellow circle). This couplet, although weak in magnitude, is indicative of a mesocyclonic circulation consistent with tornadic potential. With the MDA and its thresholds applied, the initial V$_{\text{rot}}$ at 0440 UTC was 4.07 m s$^{-1}$ with a horizontal scale of around 220 m and radar-derived vorticity of 0.073 s$^{-1}$. But, by 0450 UTC, V$_{\text{rot}}$ measurements exceeded 9.43 m s$^{-1}$ with vorticity of 0.011 s$^{-1}$ and the circulation expanded to more than 3.2 km in diameter (yellow circle at 0450 UTC). This rapid intensification and growth strongly indicate the development of a potential mesocyclone. Notably, the observed vorticity aligns well with the sounding-derived $\zeta_{\text{LLM}}$ $\sim$ 0.010$-$0.012 s$^{-1}$ lending additional confidence to the interpretation. However, because this is the first application of the MDA to this radar and the dataset carries several limitations outlined in the previous section, a strict mesocyclone classification cannot be made. Instead, we interpret the feature as a \textit{potentially mesocyclonic signature} supported by both radar and environmental evidence. Still, the location of this couplet coincides spatially and temporally with the development of a hook echo signature located at the northwestern quadrant of the supercell (Fig. \ref{fig12}a3 and Fig. \ref{fig12}a6), a classic radar feature associated with tornadic supercell. The presence of the velocity couplet and hook echo corresponds with visual and photographic evidence previously presented (Fig. \ref{fig3} and Fig. \ref{figS3}), confirming the occurrence of a tornadic mesocyclone. Throughout its evolution, the supercell propagated  east-southeastward, as depicted in  the reflectivity maps. This track is also consistent with the analyzed environmental wind profile (i.e., southwesterly low-level winds veering to westerlies and opposing upper-level northeasterlies) which collectively influenced both the propagation of the supercell and of the broader convective complex. The storm’s trajectory, along with the documented tornado damage path, is visualized in Figure \ref{fig13}. 

The vertical structure of the tornadic storm was further examined using radar vertical cross-section analysis from the S-SUB radar. As illustrated in Figure \ref{fig14}, it provides details of the vertical structure of the cell, oriented parallel to the storm motion, over the area in Candating, Arayat. A reflectivity maximum of 59 dBZ (dashed line, Fig. \ref{fig14}a1) extending beyond 6 km in altitude was observed. Complementary volume scans reveal the presence of a potential Bounded Weak Echo Region (BWER) at 0430 UTC (1230 LST; solid line, Fig. \ref{fig14}a1), an indicator of supercell storms and a signature for strong vertical motion. This signature is characterized by an area of low reflectivity surrounded by higher reflectivity aloft and laterally indicative of an extremely strong updraft that lifts hydrometeors to a significant height. This structural signature is particularly notable considering the timing of the event during the transitional month of May, when such deep convection is relatively uncommon in the region. Furthermore, the vertical reflectivity fields also suggest a tilted storm structure, consistent with the influence of environmental vertical wind shear (as previously discussed in the synoptic and mesoscale analysis). However, when interpreting radar volume scans from a single radar site such as S-SUB, it is also crucial to consider the potential for artifacts due to temporal discontinuities inherent in sequential elevation scans. These can lead to an apparent, though sometimes artificial, "pseudo-tilt” in the vertical cross-section. 

\begin{figure*}[t]
\centering
\includegraphics[width=\textwidth]{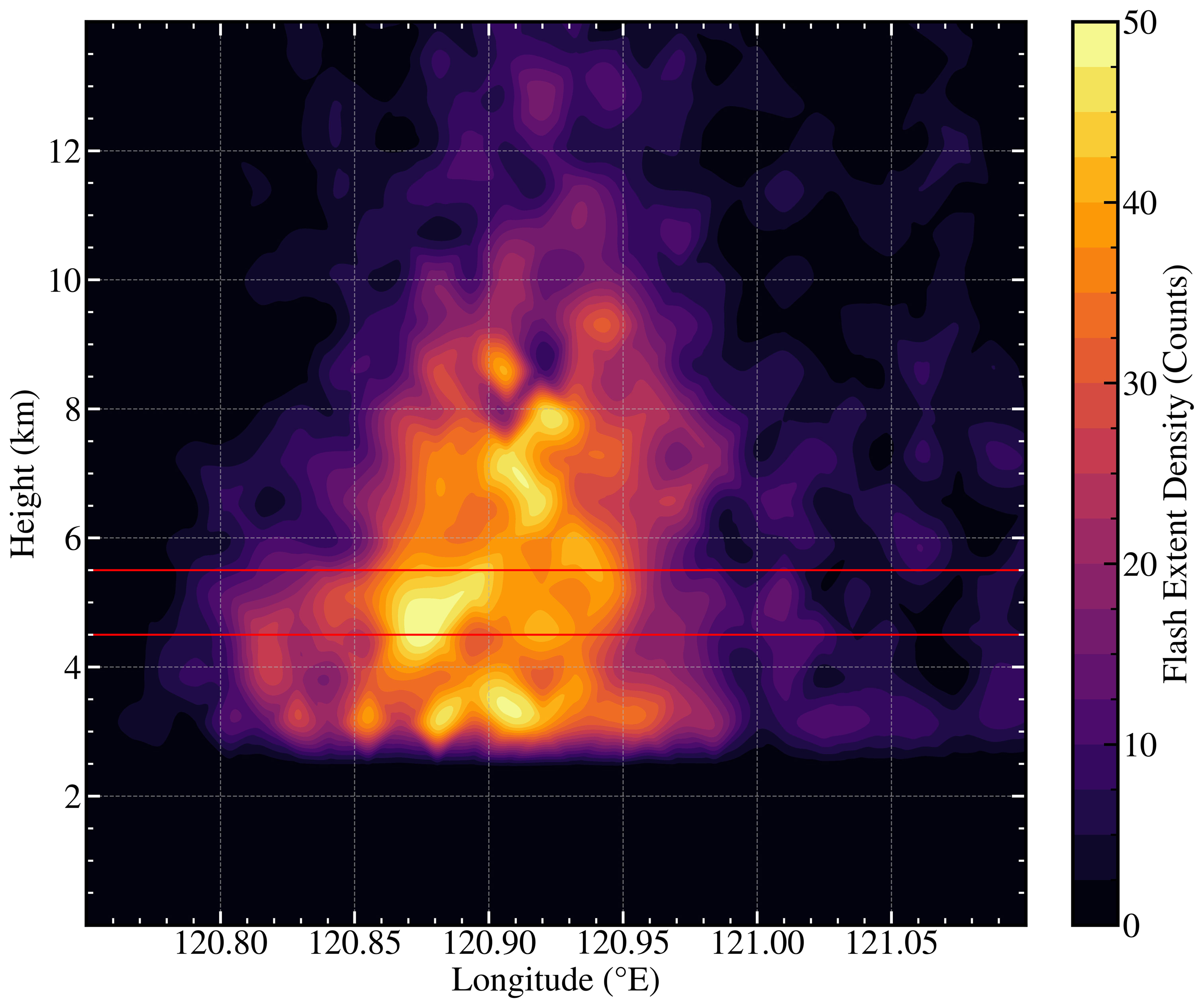}
\caption{Horizontal Cross Section of 27 May 2024 tornadic supercell’s electrical structure. Lightning observations are accumulated from 04$-$06 UTC (12$-$14 LST) and then binned in a 30$\times$30 matrix. Red lines correspond to the average freezing-level height in the tropics between 4.5$-$5.5 km AGL.}
\label{fig16}
\end{figure*}

Radial velocity cross-sections (Fig. \ref{fig14}b1-b3) further substantiate the presence of a mesocyclonic circulation. A distinct velocity couplet, marked by juxtaposed inbound (green) and outbound (red) velocities, becomes particularly evident at 0450 UTC (1250 LST), just prior to the documented time of tornadogenesis. This feature is a classical indicator of mid-level rotation and suggests the presence of a cyclonic vortex aloft. The quality control procedures applied to the radar velocity data yielded optimal clarity at the timestep, further supporting the reliability of the observed velocity signature. This pair of opposite radial velocities was located near the surface and coincided with the possible BWER and reflectivity core, affirming the intensification of the mid-level mesocyclone. This intensification temporally matched the interaction with nearby auxiliary cells $>$ 30 dBZ, which likely contributed additional inflow and low-level forcing. The subsequent enhancement of updraft strength and vorticity stretching culminated in tornadogenesis shortly thereafter. 

\subsubsection{Lightning and Microphysical Processes}

In relation to the satellite and radar analysis, Figure \ref{fig15} presents the lightning flash counts and its associated average peak currents from PLDN. As shown in Figure 15a, a peak in electrical activity occurred between 0430$-$0440 UTC (1230$-$1240 LST), preceding storm intensification indicated by decreasing brightness temperature (BT) and increasing radar reflectivity. A sharp increase in lightning activity, followed by a second lightning jump ($>$ 100 flashes min$^{-1}$; see arrows), was observed prior to the tornado occurrence at 0450 UTC (1250 LST). This pattern is in agreement with several authors \citep{Williams1989,Gatlin2010,Schultz2011,WAPLER2017}, which observed lightning activity prior to cell intensification and SWEs such as tornadoes, hail, and damaging winds, often coinciding with strong cloud-top cooling (BT $<$ $-$70 °C).

\begin{figure*}[t!]
\centering
\includegraphics[width=\textwidth]{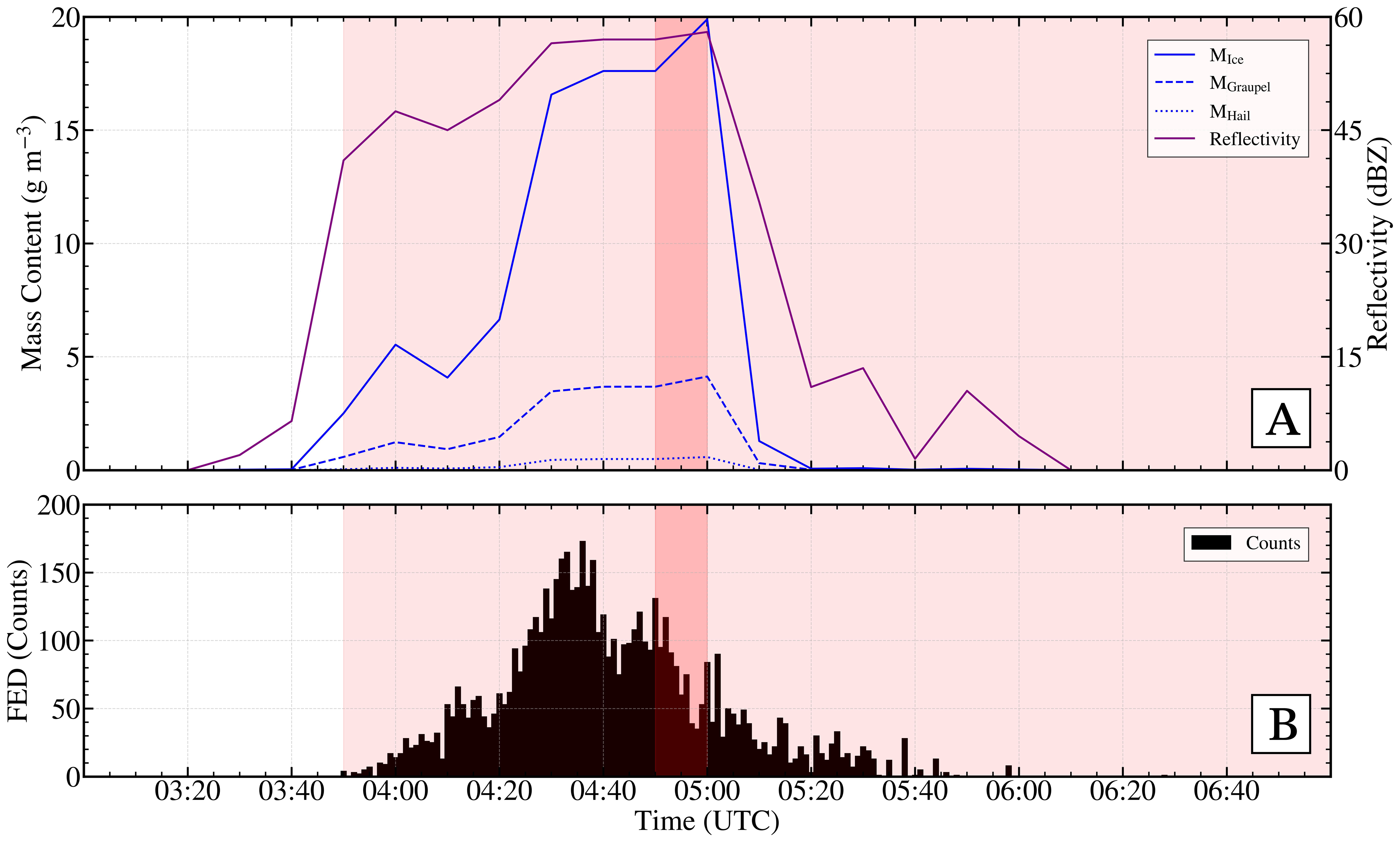}
\caption{Time series of (a) Ice (blue; solid line), Graupel (blue; dashed), Hail (blue; dotted) mass content (g m$^{-3}$) along with Reflectivity measurements (dBZ; purple; solid line) associated with the tornadic supercell. (b) Maximum flash extent density per minute (flashes min$^{-1}$). Similar red highlights are present as in Figure \ref{fig15}.}
\label{fig17}
\end{figure*}

Table \ref{table:4} summarizes the lightning characteristics associated with the Candating, Arayat tornado and its parent storm. About 10 minutes before tornadogenesis, total flash rate peaked at 173 flashes min$^{-1}$, with positive flashes dominating at 77\% (133 flashes min$^{-1}$). As shown in Figure \ref{fig15}b and detailed in Table \ref{table:4}, the overall field measurements indicate that approximately 73\% (27\%) of all lightning flashes are delivered through positive (negative) charges, with an average positive peak current of 4.3 kA (if positive bolts $>$ 10 kA are considered: 19.13 kA), similar to winter thunderstorm profiles documented by \citet{Wang2021}. These characteristics resemble those found in Spencer, South Dakota F4 tornadic supercell on 30 May 1998 \citep{Carey2003}, where positive flashes composed 77\% of total lightning, albeit with much higher peak currents exceeding 50 kA. Given the timing of the Candating tornado during the tropical transition period in May, its internal microphysical processes may differ from those of typical (JJA) convective storms. After all, lightning flash rates are often associated with enhanced rainfall rates, typical of convective storms in the Philippines during JJA, driven by intensified collision–coalescence processes among cloud hydrometeors \citep{Banares2021,MUDIAR2021,Ibanez2023}. Majority of these lightning activity $>$ 50 flashes occurred between 2$-$6 km, coinciding with the freezing layer (particularly, 4.5$-$5.5 km AGL; Fig. \ref{fig16}). This suggests a sufficient presence of ice and mixed-phase hydrometeors within the rotating updraft, possibly of smaller size and lower density. Such microphysical properties likely enhanced collisions and nucleation processes, facilitating greater charge separation and resulting in stronger and more frequent lightning discharges.

This relation between electrical activity and severe weather occurrence can be attributed to strong updrafts that support the vertical distribution of mixed-phase hydrometeors, such as graupel and ice crystals, within the storm. These updrafts, in conjunction with gravitational sorting, organize hydrometeors according to their size and altitude, enabling rapid charge separation through non-inductive electrification processes. This results in the rapid formation of positive and negative charge structures and a subsequent increase in lightning production \citep{Deierling2008}. However, the vertical positioning and intensity of charge centers can vary by season, as shown in the results of \citet{Wang2021} and \citet{Capuli2025}, likely due to differences in storm microphysics between summer and transition-season thunderstorms. Thus, identifying the type and distribution of mixed-phase hydrometeors is essential for better understanding the storm’s electrification dynamics and potential for severe weather.

The temporal evolution of total microphysical and electrical characteristics is shown in Figure \ref{fig17}. As shown in the aforementioned figure (particularly Fig. \ref{fig17}a), the mass content of key hydrometeors (ice, graupel, and hail), along with radar reflectivity, exhibited a rapid increase from the onset of the event up to their respective peaks. This intensification coincided with a marked increase in lightning flash rates, culminating in a large electrical jump between 0430 and 0450 UTC, just minutes prior to tornadogenesis (Fig. \ref{fig17}b). Following 0500 UTC, both flash rates and hydrometeor mass content began to decline, particularly the ice mass content, which peaked at $\sim$1.5$-$1.9$\times$10$^{7}$ kg (15$-$19 g m$^{-3}$). This temporal correspondence between increasing lightning activity and elevated hydrometeor concentrations aligns with the findings of \citet{Rocque2024}, who examined an SCS during the 2018 RELAMPAGO campaign. In their study, the surge in lightning activity was attributed to deep convective cores with high concentrations of cloud hydrometeor ($\geqslant$ 5.0$\times$10$^{9}$ kg), particularly near the freezing layer, reinforcing the connection between storm microphysics and electrifications

\begin{figure}[t!]
\centering
\includegraphics[width=\columnwidth]{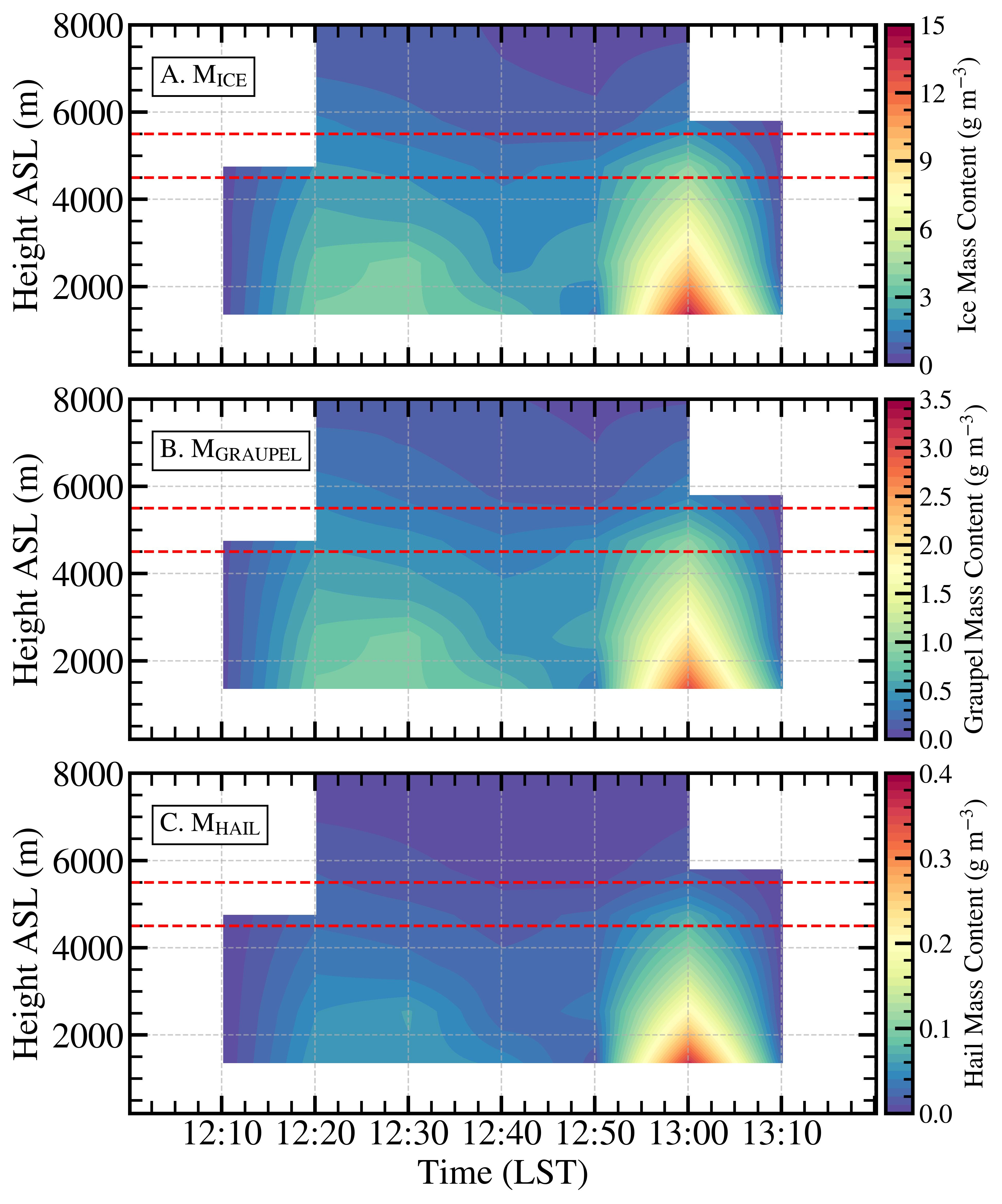}
\caption{ Time-Height Cross Sections of (a) ice mass, (b) graupel mass, and (c) hail mass content (g m$^{-3}$; shaded) of the 27 May 2024 tornadic supercell. Red lines correspond to the average freezing-level height in the tropics.}
\label{fig18}
\end{figure}

As demonstrated in previous studies \citep{Bruning2007,Bruning2015,Mecikalsi2015}, the temporal progression of lightning characteristics shown in Figures \ref{fig15}$-$\ref{fig18} reflects distinct phases of the storm life cycle. Following the framework of Bruning et al. (2007), three intervals are evident. During the initiation phase (0340$-$0400 UTC / 1140$-$1200 LST), flash rates begin to increase gradually as lightning activity becomes  more vertically oriented within the updraft, with flash centroids located below 8 km. This is followed by a modulation phase (0400$-$0420 UTC / 1200$-$1220 LST), where flash rates temporarily plateau. During this period, approximately 80\% of flashes occur below 8 km, suggesting  that charge centers are concentrated below the freezing layer due to the dominance of mixed-phase hydrometeors and frequent intracloud lightning near the storm base. The most pronounced increase in lightning flash rates occurs during the subsequent intensification phase (0420$-$0450 UTC / 1220$-$1250 LST; Fig. \ref{fig18}a-c), coinciding with the sustained growth of ice, graupel, and hail mass content just below the freezing layer and a concurrent rise in reflectivity. This elevated lightning activity persisted for about 30 minutes leading up to tornadogenesis. After 0500 UTC (1300 LST), the storm began its dissipation phase, transitioning into an outflow-dominant convective system despite nearby storm interactions.

\section{CONCLUDING REMARKS}

This case study, along with the collected observational data, provides critical insights into the storm’s evolution and surrounding environment. When integrated with archived severe weather reports from Project SWAP, the findings contribute significantly to  understanding the mechanisms that govern storm initiation and severity. Synoptic analysis reveals that the Candating, Arayat tornadic supercell developed discretely, primarily influenced by a subtle lowering of geopotential heights associated with the nearby departure of TC Ewiniar. Additional dynamical forcing was provided by a lee-side convergence mechanism and orographic influences from Mt. Arayat i.e., plain-to-mountain flow and a cyclonic lee vortex. The low-level SWM, acting as the LLJ, facilitated quality moisture advection over Central Luzon, characterized by high RH above the cloud base. Previous studies \citep[e.g.,][, and references therein]{Dial2010,Schumann2010} have shown that environments with weak synoptic-scale ascent and minimal or absent low-level linear forcing are conducive to isolated or discrete supercells with greater. Analysis of the pre-convective environment further reveals a moist atmospheric column with multiple capping inversions and minimal hodograph curvature, both of which initially inhibited morning convection. Storm initiation therefore required additional surface heating and subtle moisture injection in the low- to mid-level troposphere. 

The convective environment associated with the Candating, Arayat tornadic supercell underwent a notable evolution in its thermodynamic and kinematic profile, as indicated by the Skew-T Hodograph during the event timeframe. Initially, at 00 UTC (08 LST), the profile exhibited instability just above the PBL. By 04$-$06 UTC (12$-$14 LST), however, the profile transitioned into an $inverted-V$ structure characterized by steep low-level lapse rates, particularly within the first 3 km. This supported the buildup of undiluted CAPE $>$ 3000 J kg$^{-1}$, which was highly favorable for convective initiation. Although the LLS vector and storm-relative inflow remained weak throughout the period, several environmental factors likely contributed to the eventual tornadogenesis. These include; (i) sufficient 0-6 km bulk shear ($\sim$20 kt), supported by modest curvature in the storm-relative hodograph, (ii) localized lee-side convergence, which may have enhanced the LLS and background vorticity by modifying the low-level southwesterlies and northwesterlies as they channelled over elevated terrain and converged downstream, and (iii) interactions with nearby external storm interactions through the presence of auxiliary cells, which may have further modulated and balanced the storm’s inflow. Additionally, terrain-induced effects likely contributed to increased streamwiseness of the horizontal vorticity compared to the morning profile, suggesting that the storm’s updraft was better positioned to ingest, stretch, and tilt pre-existing vorticity into vertical vorticity within the developing low-level mesocyclone. Collectively, these conditions favored the development of a stout updraft capable of supporting tornadogenesis as the storm propagated away from the higher terrain.

The integration of multiple meteorological data sources, including netizen reports, lightning data, satellite data, and single-polarization S-band radar data from S-SUB, confirms that the tornado occurred during the mature phase of the parent storm’s life cycle 0450 UTC (1250 LST), and persisted for approximately 10 minutes. This event coincided with the presence of cloud overshooting tops and a $\sim$2 km long damage path,  as captured by Landsat-9 OLI. Satellite and radar analyses further reveal a concurrent increase in radar reflectivity ($>$ 50 dBZ), alongside the development of auxiliary convective cells in the storm’s initiation region, specifically along the rear and left flanks, which likely contributed to storm intensification and subsequent tornadogenesis \citep{Nixon2024}.Key radar-based signatures, including a possible BWER signature, velocity couplet, and the identification of mesocyclone using MDA parameters and its thresholds, provide additional confirmation of tornado occurrence. These features were further supported by a notable surge in lightning activity ($>$ 100 flashes min$^{-1}$), several minutes prior to the SWE occurred. The increased lightning rates, occurring below the freezing layer, are attributed to the active presence of cloud hydrometeors, such as ice, graupel, and hail, undergoing microphysical processes known to enhance storm electrification and intensify convective updrafts \citep{Deierling2008}. 

However, the absence of a discernable precipitation curtain in this supercell remains unclear to us (the authors), potentially hinting a Low-Precipitation (LP) supercell mode, despite the presence and superposition of cloud hydrometeors. Typically, higher lightning flash rates are associated with higher rainfall intensities due to their influence on the collision-coalescence processes among hydrometeors \citep{MUDIAR2021}. One hypothesis is that the weak LLS did not significantly interact with the storm’s strong updraft, thus modifying neither its dynamics nor its microphysics, unlike other supercell precipitation modes which often involve larger concentrations of cloud condensation nuclei (CCN) and ice nuclei \citep{Proctor1983,Weisman1985}. Alternatively, precipitation could have been displaced from the core by upper-level storm-relative flow, as proposed by \citet{RasmussenStraka1998}. However, this appears contradictory, as LLS may have been enhanced through terrain-induced effects at the storm’s initiation area as previously discussed. Another is the presence of the BWER, which signifies an intense updraft, is potentially capable of preventing or severely limiting precipitation from falling through the storm’s core at lower-to-mid levels. Furthermore, the supercell eventually evolved into a mesoscale convective system (MCS) during its latter stage. \citet{James2010} emphasized that mid-level dryness influences deep convection in ways that are highly sensitive to the prevailing microphysical processes. In this context, our findings align more closely with the work of \citet{Grant2014}, which demonstrated that moisture content just above a well-mixed boundary layer plays a critical role in determining supercellular mode and the distribution of cloud hydrometeors, a conclusion also supported by the aforementioned studies and illustrated in the ERA5 Skew-T Profiles (Fig. 8). These  findings also support the recent study by \citet{Yuan2024} which examined a tornadic storm in Jiangsu Province, China on 14 May 2021, and similarly found distinct  microphysical characteristics between tornadic and non-tornadic storms.

This study aimed to; (i) characterize the tornadic supercell features of the Candating, Arayat storm, (ii) emphasize the environmental complexities, particularly mesoscale processes, during the tornado episode, as they may have substantially influenced tornadogenesis, and (iii) utilize remote sensing techniques to delineate the SWE. In the Philippines, ingredient-based analyses of severe storm environments often rely on convective-kinematic parameters derived from North American and European severe weather climatologies. This allowed for comparisons, as well as identification of discrepancies, between Philippines tornado environments and more well-documented ‘Great Plains setups’, particularly in light of mesoscale inhomogeneities and the uncertainties inherent to using operational or reanalysis datasets for such investigations. It is hoped that the insights and points presented in this study will raise awareness of Filipino weather forecasters and meteorologists regarding both the importance and the challenges of recognizing mesoscale factors that can either enhance or suppress  tornado potential (or any SWE). In the longer term, this work may encourage more systematic studies on tornadic storms in the country. This case study is, to our knowledge, the first to analyze a tornadic supercell in the Philippines, using numerous observational data and sounding parameters presented here. Thus, it serves as a valuable foundation from which more extensive studies can build upon.

Our experience also suggests that the case study presented here is broadly representative of the severe storms that recurrently affect Central Luzon and adjacent areas, including the National Capital Region (NCR), Region IV$-$A (CALABARZON). These storms, like the one examined in this work, frequently reach severe intensity. Recent findings from Project SWAP \citep{Capuli2024b} indicate that the Philippine tornado season generally begins in May and clears off between September and October. The tornadic supercell that impacted Candating, Arayat occurred at the onset of this severe weather season and within one of the identified tornado ‘hotspot’ regions. This underscores the need to develop a climatology of severe weather environments that does not depend solely  on rainfall, temperature, or other near-surface parameters, but instead focuses on instability and kinematic profiles throughout the atmospheric  column and/or within specific layers. Such an approach is essential for more accurately diagnosing cases associated with classical southwesterly flow patterns, such as the tornadic event documented here.

Finally, certain limitations of this study must be acknowledged. One key limitation lies in the damage analysis, which, though guided by standardized methodologies, remains an informed approximation rather than an exact  measurement. The damage analysis was conducted without formal structural engineering expertise and did not account for key structural variables such as construction quality, building age, or maintenance history, all of which can significantly influence the degree of observed damage. Moreover, the EF scale was developed under the United States with specific structural standards and engineering assumptions. Thus, its direct application in the Philippine context may lead to over- or underestimation of actual wind speeds due to differences in construction practices and building resilience. Despite these limitations, the EF scale was employed here as a scientifically validated and internationally recognized method for estimating wind speeds based on DIs. To improve future assessments, case studies in the Philippines are encouraged to adopt or reference the International Fujita (IF) scale, developed by European Severe Storms Laboratory \citep[ESSL;][]{Groenemeijer2023}. Designed to account for variability in building types and structural integrity across different countries, the IF scale provides a more globally adaptable framework for tornado damage assessment and may yield more accurate, context-sensitive estimates in regions like the Philippines.

Remote sensing constraints also influenced the spatial assessment of damage. Although Landsat-9 OLI with NDVI analysis, enhanced through pansharpening techniques, yielded valuable insights into post-event conditions, its native 30-m spatial resolution remains a limiting factor. This resolution may obscure finer-scale damage patterns, particularly in heterogeneous or fragmented landscapes. To improve spatial fidelity, succeeding studies should consider incorporating higher-resolution satellite imagery if available (e.g., Sentinel-2, PlanetScope) or acquire drone-based remote sensing to more precisely detect and quantify vegetation and surface along the damage path. \citet{Pucik2024} demonstrated the effectiveness of such an approach, combining the IF scale with Sentinel-2 10-m imagery, and citizen reports to delineate a violent tornado that occurred in Czechia back on 24 June 2021, which was rated IF4 based on damage analysis.  

Another key limitation is the lack of high-resolution numerical modeling, such as the use of Weather Research and Forecasting (WRF) or Cloud Model 1 (CM1) model, which could have provided additional insights into the complex processes within the tornadic environment. This study instead relied on available resources, including ERA5 reanalysis data, to assess the large-scale atmospheric conditions. While ERA5 provides valuable meteorological context, it cannot resolve storm-scale convection or other mesoscale processes in the way that a cloud-resolving model like WRF or CM1 can. For example, \citet{Scheffknecht2017} used high-resolution modeling to examine the life cycle of a supercell in the northern side of the Alpine mountain range, revealing how southwesterly synoptic flow combined with thermally-induced plain-to-mountain flow near the surface created favorable conditions for a supercell initiation. Such detail lies beyond the scope of ERA5, yet reanalysis datasets remain reliable for capturing the broader synoptic and mesoscale environments. In this study, ERA5 was essential for characterizing the pre-, during, and post-storm conditions of the event (where the SWE occurred around 05 UTC / 13 LST), consistent with other severe weather case studies utilizing similar reanalysis products \citep[e.g.,][]{Oliveira2022}. Nevertheless, future research would benefit from high-resolution simulations to better resolve tornadic storm dynamics and assess the influence of topography on storm structure and intensity.

Finally, this study makes use of the available radar data. The inclusion of radar analysis is commendable, especially given that many past severe weather case studies in the Philippines lacked such observations. However, the analysis is severely constrained by the limitations of a single-polarization radar, such as the S-SUB radar, including the de-aliasing process in this study given that this is the first time that a quality control procedure for radial velocity field was introduced and tackled in the radar system of DOST-PAGASA and the usage of thresholds for MDA. In fact, the low unambiguous limit for the radial velocity product, likely due to instrumental degradation or issue, of S-SUB is another major factor to this limitation as the radar can misinterpret high inbound/outbound velocities as having the opposite sign (e.g., a $+$25 m s$^{-1}$ wind appears as $-$5 m s$^{-1}$). The folded values make it difficult to estimate the ‘true’ rotational velocity, often leading to underestimation. Furthermore, when too much of the field is aliased, the de-aliasing algorithm may select the wrong phase correction, introducing systematic biases or patchy artifacts in the de-aliased field which can lead to reduced confidence in identifying or quantifying small, intense circulations. Given these limitations in the radar analysis, and so as to its products, it is a must to develop a cohesive quality control procedure of radar products tailored to forecasting and analysis of SCSs 
\citep[e.g.,][]{Veillette2023}. 

Dual-pol radars (whether C-, S- or X-band) can also provide more detailed and critical insights into storm structure, reflectivity patterns, and hydrometeor classification, which are essential for accurately assessing storm intensity and evolution. For example, \citet{Yuan2024} demonstrated that analyzing Z$_{\text{DR}}$, K$_{\text{DP}}$, and drop size distribution (DSD) trends can help distinguish microphysical contrasts between tornadic and non-tornadic supercells, aiding both prediction and warning efforts. Dual-polarization radars also complement lightning observations, improving understanding of the electrical structure of SCSs \citep[e.g.,][and references therein]{Rocque2024}. Not only do they complement such observations, but the dual-pol moments are also seen to be used in more advanced detection of supercell structures such as mesocyclones and other features \citep[e.g., Supercell Polarimetric Observation Research Kit/SPORK;][]{VanDenBroeke2022}. Unfortunately, due to beam-blockage from the nearest dual-polarization S-band radar in Baler, Aurora, such data were unavailable for this case. Additionally, radar sampling resolution degrades with distance as averaging occurs across a wider beam, smoothing extreme reflectivity and/or velocity values, and potentially weakening or obscuring signatures such as BWERs, mesocyclones, or velocity couplets. Given these constraints, the present analysis relied on other resources, including ground-based observations, synoptic data, and satellite imagery, to characterize the SWE.

Moving forward, initiatives like DOST-PAGASA’s Radar Application on Data Assimilation and Rainfall NowCASTing (RADARCAST), part of the agency’s  collaborative effort with Taiwan’s Central Weather Administration and other partners, offer promising opportunities. By enhancing radar data quality and assimilating both single- and dual-pol products (and eventually, the transition from using single-pol to dual-pol radars), these efforts could yield higher-quality observations near SCSs. Despite these limitations, this study demonstrates that valuable and comprehensive case documentation of SWEs, particularly tornadoes, can be achieved using limited resources. Future investigations should prioritize integrating high-resolution numerical modelling with expanded and radar coverage, including upgrades and new radars (such as the Bataan Radar/S-BAT, that will likely replace S-SUB) to the Philippine Radar Network, to improve the characterization of tornadic and other SWEs in the country.

\acknowledgments

We are very grateful to the Philippine population for reporting the occurrence of the tornado in the community. G. H. Capuli, M. A. O. Noveno, and M. P. A. Ibañez also appreciated the valuable comments of the three anonymous reviewers and editor, which helped to improve this manuscript. This work received no funding, but was ’funded’ by extensive and exhaustive effort, whose dedication and commitment to advancing our understanding of severe weather phenomena were indispensable. We are thankful to our respective family and loved ones for their unwavering support throughout this research.

\contribution

\textbf{Generich H. Capuli:} Writing - original draft, Writing - review \& editing, Visualization, Methodology, Investigation, Formal Analysis, Data curation, Conceptualization, Supervision. \textbf{Michael Angelo O. Noveno:} Writing - review \& editing, Visualization, Methodology, Investigation, Formal Analysis, Data curation. \textbf{Marco Polo A. Ibañez:} Writing - review \& editing, Visualization, Methodology, Investigation, Formal Analysis, Data curation

%
%
\datastatement

Data used in this paper were derived from the ERA5 reanalysis (openly available through \href{https://cds.climate.copernicus.eu/#!/home}{Climate Data Source}), the individual full-disk HIMAWARI-8/9 Bands were accessible through \href{https://thredds.nci.org.au/thredds/catalog/catalogs/ra22/satellite-products/arc/obs/himawari-ahi/himawari-ahi.html}{THREDDS Catalog}. The 2023 Philippines Administrative Level 0-4 shapefiles are available in Humanitarian Data Exchange (\href{https://data.humdata.org/dataset/cod-ab-phl}{HDX}). The author’s associated 1D vertical profile of standard ERA5 data and sounding data will be available soon through Project SWAP, but can be replicated using the location data through SounderPy (v3.0.8) at the lat-lon coordinates of (15.15 °N, 120.80 °E). The Digital Elevation Model (DEM) is from Copernicus GLO-30 Digital Elevation Model distributed and available in \href{https://doi.org/10.5069/G9028PQB}{OpenTopography}. Finally, the lightning data from the PLDN and radar data are requested through DOST-PAGASA, but can also be obtained from the corresponding author upon reasonable request. Proper attribution is required for these datasets.
\\
\\
This paper has made of use of the following Python packages:  \verb|Cartopy|, \verb|GeoPandas|, \verb|Matplotlib|, \verb|MetPy|, \verb|NumPy|, \verb|Pandas|, \verb|Rasterio|, \verb|rioxarray|, \verb|SciPy|, \verb|SounderPy|, and \verb|xarray|.

\interest

The authors declare that they have no competing interests.

%

\appendix




\appendixtitle{Supplementary Figures}

\begin{figure*}[t]
\centering
\includegraphics[width=\textwidth]{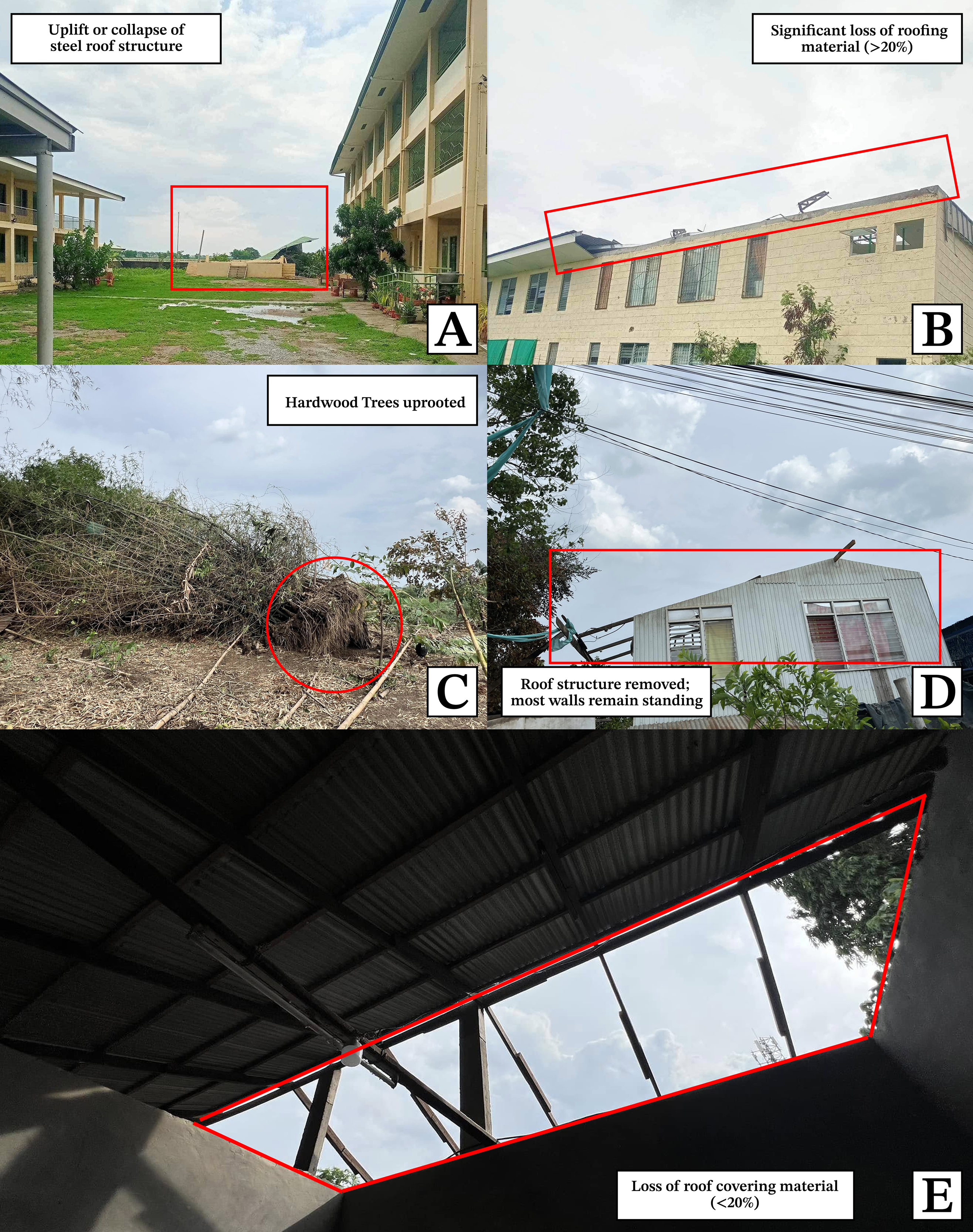}
\caption{Photographs of the damage extent brought by the Arayat, Candating Tornado. Photographs are taken by Konsi Jace.}
\label{figS1}
\end{figure*}

\begin{figure*}[t]
\centering
\includegraphics[width=\textwidth]{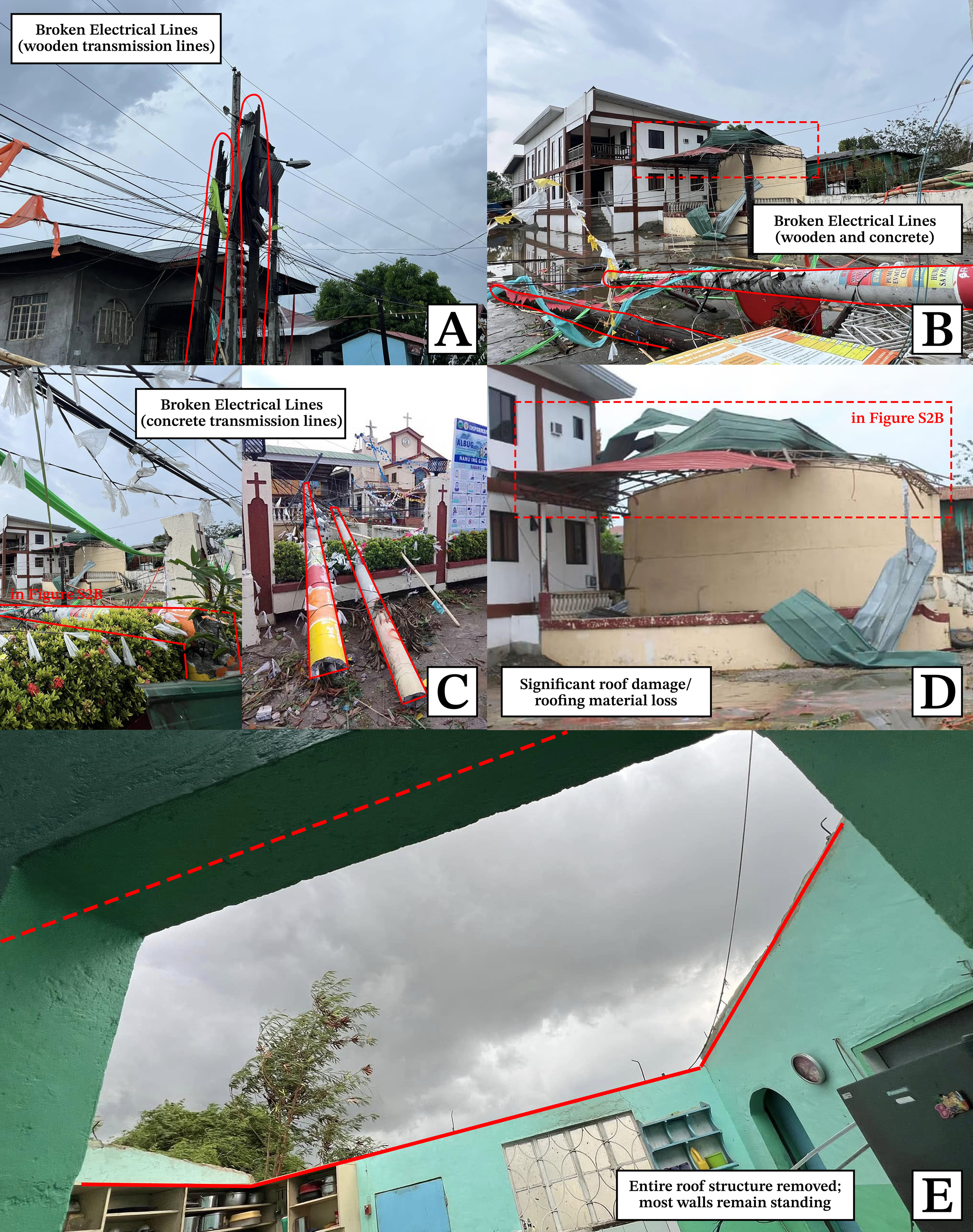}
\caption{Similar to Figure \ref{figS1}, additional photographs of the damage extent brought by the Arayat, Candating Tornado. Photographs are taken by Konsi Jace.}
\label{figS2}
\end{figure*}

\begin{figure*}[t]
\centering
\includegraphics[width=\textwidth]{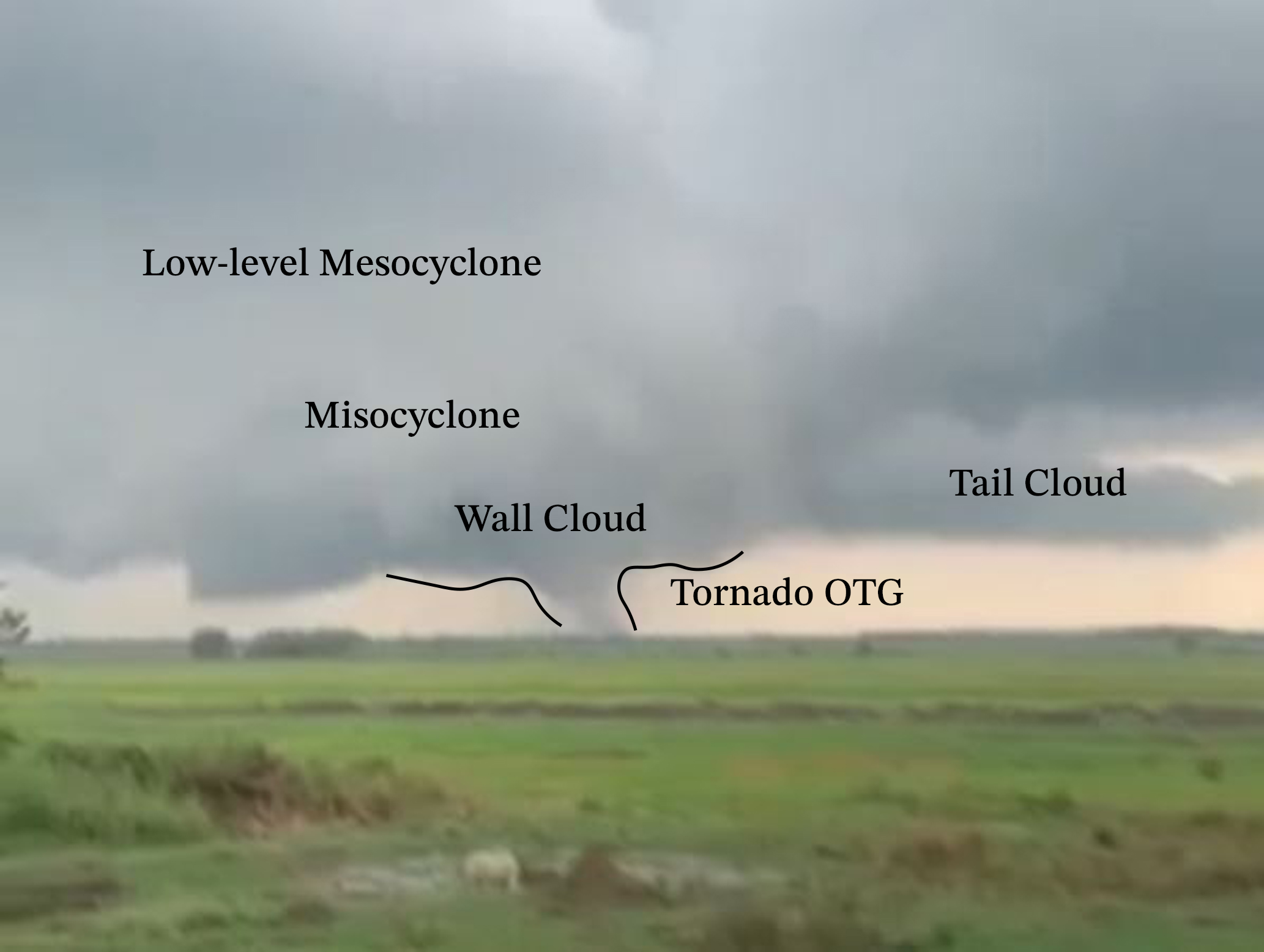}
\caption{Additional photograph of the tornadic storm from a distance. Annotations to the structures and features identified were included. Video captured by Willie Martin.}
\label{figS3}
\end{figure*}

Attached are Figures \ref{figS1}, \ref{figS2}, and \ref{figS3}.

%




\bibliographystyle{ametsocV6}
\bibliography{references}

\end{document}